\documentclass[12pt]{article}
\usepackage[latin1]{inputenc}

\usepackage{amsmath}
\usepackage{amsfonts}
\usepackage{amssymb}
\usepackage{graphicx}
\usepackage{geometry}
\usepackage{amssymb,epsfig,subfigure}
\usepackage{hyperref}

\makeatletter
\renewcommand\section{\@startsection {section}{1}{\z@}%
                                 {-3.5ex \@plus -1ex \@minus -.2ex}
                                   {2.3ex \@plus.2ex}%
                                   {\normalfont\large\bfseries}}
\renewcommand\subsection{\@startsection{subsection}{2}{\z@}%
                                   {-3.25ex\@plus -1ex \@minus -.2ex}%
                                     {1.5ex \@plus .2ex}%
                                     {\normalfont\bfseries}}
\renewcommand\subsubsection{\@startsection{subsubsection}{3}{\z@}%
                                   {-3.25ex\@plus -1ex \@minus -.2ex}%
                                     {1.5ex \@plus .2ex}%
                                     {\normalfont\itshape}}
\makeatother

\def\pplogo{\vbox{\kern-\headheight\kern -29pt
\halign{##&##\hfil\cr&{\ppnumber}\cr\rule{0pt}{2.5ex}&\ppdate\cr}}}
\makeatletter
\def\ps@firstpage{\ps@empty \def\@oddhead{\hss\pplogo}%
  \let\@evenhead\@oddhead 
}
\def\maketitle{\par
 \begingroup
 \def\thefootnote{\fnsymbol{footnote}}
 \def\@makefnmark{\hbox{$^{\@thefnmark}$\hss}}
 \if@twocolumn
 \twocolumn[\@maketitle]
 \else \newpage
 \global\@topnum\z@ \@maketitle \fi\thispagestyle{firstpage}\@thanks
 \endgroup
 \setcounter{footnote}{0}
 \let\maketitle\relax
 \let\@maketitle\relax
 \gdef\@thanks{}\gdef\@author{}\gdef\@title{}\let\thanks\relax}
\makeatother

\numberwithin{equation}{section}

\newcommand{\be}{\begin{eqnarray}}
\newcommand{\bea}{\begin{eqnarray}}
\newcommand{\ee}{\end{eqnarray}}
\newcommand{\eea}{\end{eqnarray}}
\newcommand{\beq}{\begin{equation}}
\newcommand{\eeq}{\end{equation}}

\newcommand{\f}{\frac}
\newcommand{\mc}{\mathcal}
\newcommand{\Tr}{{\rm Tr}}
\newcommand{\tr}{{\rm tr}}
\renewcommand{\t}{\tilde}
\newcommand{\muphi}{\mu_\phi}

\begin{document}

\setcounter{page}0
\def\ppnumber{\vbox{\baselineskip14pt
}}
\def\ppdate{\footnotesize{\tt SLAC-PUB-14247}} \date{}

\author{Siavosh R. Behbahani$^{a,b}$, Nathaniel Craig$^{b,c,d}$,  Gonzalo Torroba$^{a,b}$\\
[7mm]
{\normalsize \it $^{a}$  SLAC National Accelerator Laboratory, Stanford, CA 94309}\\
{\normalsize \it $^{b}$ Department of Physics, Stanford University, Stanford, CA 94305}\\
{\normalsize \it $^{c}$ Institute for Advanced Study, Princeton, NJ 08540}\\
{\normalsize \it $^{d}$ Department of Physics and Astronomy, Rutgers University, Piscataway, NJ 08854}\\
[3mm]
}

\bigskip
\title{\bf  Single-sector supersymmetry breaking, chirality and unification
\vskip 0.5cm}
\maketitle


\begin{abstract} \normalsize
\noindent
Calculable single-sector models provide an elegant framework for generating the flavor textures via compositeness, breaking supersymmetry, and explaining the electroweak scale. Such models may be realized naturally in supersymmetric QCD with additional gauge singlets (SSQCD), though it remains challenging to construct models without a surfeit of light exotic states where the Standard Model index emerges naturally.  We classify possible single-sector models based on Sp confining SSQCD according to their Standard Model index and number of composite messengers. This leads to simple, calculable models that spontaneously break supersymmetry, reproduce the fermion flavor hierarchy, and explain the Standard Model index dynamically with little or no additional matter. At low energies these theories realize a ``more minimal'' soft spectrum  with direct mediation and a gravitino LSP. 
\end{abstract}
\bigskip
\newpage

\tableofcontents

\vskip 1cm

\section{Introduction}\label{sec:intro}

A very appealing idea is that some of the Standard Model (SM) quarks and leptons are secretly composites of more fundamental ``preons'' that explain their attributes in a simple way. Seemingly unrelated puzzles like the origin of the Yukawa couplings and the gauge hierarchy problem could then be solved by the same underlying dynamical mechanism. This more fundamental gauge theory has to produce approximately massless fermionic bound states, protected by a chiral symmetry.\footnote{These ideas have been explored over the years in different ways, mainly motivated by the work of 't Hooft~\cite{'tHooft:1979bh}. Early examples of nonsupersymmetric models were given in~\cite{nonsusy}; for a review and references on supersymmetric constructions see~\cite{Volkas:1987cd}. A very interesting model combining technicolor with composite SM fermions was given in~\cite{Kaplan:1991dc}.}
Supersymmetric gauge theories provide a natural framework for these ideas. Indeed, supersymmetric QCD (SQCD) can produce exactly massless bound states. It may also be used to generate dynamically the Fermi scale and to break supersymmetry. Furthermore, Seiberg duality~\cite{Seiberg:1994pq} gives, in many cases, weakly coupled dual theories where the massless mesons and baryons are described as elementary excitations.

Recent developments in SUSY gauge theories have motivated a renewed interest in constructing realistic composite models. The works~\cite{ArkaniHamed:1997fq, Luty:1998vr} proposed that the strong dynamics responsible for supersymmetry breaking may also produce composite SM fermions. These ``single-sector models'' can in principle give a simultaneous explanation for the observed flavor textures and the stabilization of the electroweak hierarchy. They are also quite economical in that they do not have a modular structure with messengers put in by hand --supersymmetry is communicated directly to the composites. These beautiful constructions have the drawback of not being calculable, and a detailed understanding of the spectrum was not possible.

The insight of Franco and Kachru~\cite{Franco:2009wf} was to combine these ideas with the ISS mechanism (Intriligator, Seiberg and Shih~\cite{Intriligator:2006dd}), building calculable single-sector models in SQCD. Next~\cite{Craig:2009hf} constructed models with a fully realistic texture using a dimensional hierarchy mechanism. The flavor hierarchies are generated by coupling the SQCD mesons to elementary Higgs fields via higher dimensional operators, produced at a certain scale $M_{flavor}$ larger than the dynamical scale $\Lambda$. After confinement, these irrelevant operators give rise to marginal Yukawa interactions, naturally suppressed by powers of
\begin{equation}\label{eq:intro1}
\epsilon = \Lambda/ M_{flavor}\,.
\end{equation}
Moreover, following the same single sector philosophy it was shown in~\cite{SchaferNameki:2010iz} that the strong gauge dynamics can also yield a composite Higgs, break $SU(2) \times U(1)$ and solve the $\mu/B_\mu$ problem.

The main conclusion from these works is that SQCD in the free magnetic range, supplemented by adequate superpotential deformations, can provide a unified explanation for the flavor hierarchies, supersymmetry breaking and Higgs physics, simultaneously generating some of the SM fields as composites. The requirement of being in the free magnetic range is essential for calculability: the asymptotically free ``electric theory'' becomes strong in the IR but admits a weakly coupled ``magnetic dual''. The magnetic theory gives a description of the SM composites as elementary excitations, allowing for a direct perturbative analysis of their interactions and spectrum.

\vskip 2mm

In this context, there are two important aspects that have to be addressed. The first one is how the SM chirality emerges from the confining gauge theory. Is it possible to generate dynamically a nonzero index\footnote{The index for a complex representation $\bf R$ is defined as $N_{\bf R} - N_{\overline{\bf R}}$.} for the ${\bf 10}+ {\bf \bar 5}$ representations of $SU(5)_{SM}$? Is there a simple explanation for the family structure of the SM? The other point regards the construction of single sector models with perturbative unification. This has been quite difficult to achieve in constructions so far. The goal of this work is to address these points.

\subsection{Overview}\label{subsec:overview}

Let us summarize the main points of the present work. In \S \ref{sec:chirality} a general analysis of calculable single-sector models is presented, with emphasis on SM chirality and possible flavor textures. In models constructed to date, the confining interactions generate, for each desired ${\bf 10}+ {\bf \bar 5}$ composite generation, unwanted matter in the conjugate ${\bf \overline{ 10}}+ {\bf 5}$. The problem of conjugate representations was avoided in~\cite{Franco:2009wf,Craig:2009hf} by introducing additional spectator fields in ${\bf 10}+ {\bf \bar 5}$ multiplets, and coupling them to the unwanted matter. The procedure is somewhat artificial, because the ${\bf 10}+ {\bf \bar 5}$ index is put in by hand directly in the UV. This is overcome by allowing the electric quarks to transform under chiral representations of the SM gauge group.

The problems of generating the SM chirality and achieving unification are related: models with chiral SM representations have in general less amount of unwanted composites than their vector-like counterparts. This helps to keep the gauge couplings perturbative. Motivated by these two points, we perform in \S \ref{sec:general} a general classification of confining SQCD theories that produce composites in chiral representations of the SM gauge group. This group theory analysis will be used to construct more efficient single sector models with less amount of extra matter and spectators. 

This reveals a rich set of possible spectators (denoted by $S$), both in chiral and vector-like representations of $SU(5)_{SM}$.
The supersymmetry breaking structure of single-sector models is based on the ISS mechanism~\cite{Intriligator:2006dd}, but it includes an important novelty: some of the mesons are coupled to the magnetic singlets $S$. In the electric theory, these couplings are generated from marginal interactions between the electric quarks $Q_i$ and spectators,
\begin{equation}\label{eq:couplingSQQ}
W_{el} = \lambda\,S_{ij} Q_i Q_j\,.
\end{equation}
These become relevant in the IR. In \S \ref{sec:metastable} we present a detailed analysis of metastable supersymmetry breaking in the presence of the deformations (\ref{eq:couplingSQQ}) for $S$ in real or complex representations of the flavor group.

In \S \S \ref{sec:SU2} and \ref{sec:dimh} we apply our results to construct
examples where a nonzero ${\bf 10}+ {\bf \bar 5}$ index is dynamically generated, and which perturbatively unify. The first two composite generations can either arise from a single electric meson (in which case there is an approximate $U(2)$ flavor symmetry) or from mesons of different classical dimension (``dimensional hierarchy'' models). The third generation and $H_u$ are necessarily elementary.
An interesting outcome of our analysis is that the more economical examples are in fact models with $U(2)$ flavor symmetry. However, the flavor textures generated via Eq.~(\ref{eq:intro1}) are not in full agreement with experimental values. We will present a new mechanism for generating fermion masses and mixings that circumvents this problem. 

Insofar as the calculability of our models relies on preserving perturbativity of SM gauge couplings up to the GUT scale, we briefly review perturbativity constraints on additional SM-charged matter in Appendix \ref{sec:unif}. Finally, we end in Appendix \ref{sec:FCNC} with a detailed discussion of FCNC constraints on the pattern of soft supersymmetry-breaking masses in the various models under consideration. While $U(2)$-symmetric models are essentially unconstrained by FCNCs, mild constraints arise for models with a dimensional hierarchy.

\section{Chirality and flavor hierarchies}\label{sec:chirality}

We begin by describing our approach in general terms; some of the results can be applied to mechanisms different than the ones presented in~\cite{Franco:2009wf,Craig:2009hf,SchaferNameki:2010iz}. In order to clarify our motivations and the classification given in \S \ref{sec:general}, some necessary results from these papers will also be summarized in this section.

In searching for microscopic gauge theories that can produce a composite SM it is useful to review the way in which the SM itself explains the observed hadrons. In the IR, $SU(3)_C$ becomes strong and confines, while the $SU(2) \times U(1)$ gauge fields are `spectators' of the strong color dynamics. The SM also contains spectator fermions that only couple to $SU(2) \times U(1)$. These are of course the leptons and, in particular, they are crucial for anomaly cancellation. The theory of preons that could underlie the SM will be built upon a similar pattern. This approach was advocated for instance in~\cite{'tHooft:1979bh}.

\subsection{Basic SQCD setup}\label{subsec:basics}

For concreteness we will consider a SQCD theory with $Sp$ gauge group. The analysis for $SU$ and $SO$ theories is quite similar, and it turns out that $Sp$ theories lead to more economical models. Recall that an $Sp(2N_c)$ gauge theory\footnote{The notation is $Sp(2) \sim SU(2)$ and, more generally, $Sp(2N) \subset SU(2N)$.} with $2N_f$ fundamental quarks $Q_i$, $i=1,\ldots,2N_f$ ($N_f$ flavors) admits, for $N_f>N_c-2$, a dual description with gauge group
$$
Sp\left(2 \t N_c \equiv 2 (N_f-N_c-2)\right)
$$
containing $2N_f$ magnetic quarks $q_i$ together with a meson singlet $M_{ij} = Q_iQ_j$ in the antisymmetric of the flavor group. The dynamical scale of the theory is denoted by $\Lambda$. Following the discussion in \S \ref{sec:intro} we require that the magnetic dual is IR free, $N_c+3 \le N_f < 3(N_c+1)/2$.

In some cases we will also be interested in adding a field $U$ in the ``traceless'' antisymmetric ($N_c(2N_c-1)-1$) of the gauge group, with a superpotential $W \propto \Tr (J_{2N_c}U)^3$ which restricts the mesons to
\begin{equation}\label{eq:defM}
M_{ij} \equiv Q_i  Q_j\;,\;(M_{U})_{ij} \equiv Q_i U Q_j\,.
\end{equation}
Both are in the antisymmetric of the flavor group $SU(2N_f)$. Color indices are contracted with $J_{2N_c}= \mathbf 1_{N_c} \otimes (i \sigma_2)$. The motivation for adding this field is to have mesons of different classical dimension that can lead to realistic fermion masses~\cite{Craig:2009hf}. The duality for this case was studied in~\cite{Intriligator:1995ff}.

The SM quantum numbers are explained by weakly gauging $SU(5)_{SM} \subset SU(2N_f)$ and taking the electric quarks to transform under (possibly) chiral representations ${\bf R}_i$ of $SU(5)_{SM}$. Finally, the theory contains elementary spectator fields $S_a$ which, by definition, are singlets under $Sp(2N_c)$ and transform under $SU(5)_{SM}$. In summary, the matter content is given by
\begin{center}
\begin{tabular}{c|cc}
&$Sp(2N_c)$&$SU(5)_{SM}$\\
\hline
$Q_i^\alpha$& $\Box$ & ${\bf R}_i$  \nonumber\\
$U_{\alpha \beta}$& antisym & $\bf 1$  \nonumber\\
$S_a$& $\bf 1$ & ${\bf R}_a$
\end{tabular}
\end{center}
(By a slight abuse of notation, the same indices $i$, $j$ are used to denote the $N_f$ flavors $Q_i$ and the possible representations ${\bf R}_i$). According to our definition, all the elementary SM fields (third generation, Higgs, etc.) are contained in the spectator fields $S_a$. Anomaly cancellation may also require extra elementary spectators not present in the SM. 

The structure of the magnetic dual is
\begin{center}
\begin{tabular}{c|cc}
&$Sp(2 \t N_c)$&$SU(5)_{SM}$\\[3pt]
\hline
&&\\[-12pt]
$M_{ij}$& $\bf 1$ & ${\bf R}_{[i} \times \overline {\bf R}_{j]}$  \nonumber\\
$(M_U)_{ij}$& $\bf 1$ & ${\bf R}_{[i} \times \overline {\bf R}_{j]}$  \nonumber\\
$q_i^\alpha$& $\Box$ & $\overline {\bf R}_i$  \nonumber\\
$\t U_{\alpha \beta}$& antisym & $\bf 1$  \nonumber\\
$S_a$& $\bf 1$ & ${\bf R}_a$
\end{tabular}
\end{center}
where $\t N_c= N_f - N_c -2$ in the absence of $\t U$, while including this field with a cubic superpotential leads to $\t N_c=2(N_f-2)-N_c$. The SM quantum numbers of the mesons are obtained from the antisymmetric part of ${\bf R}_i \times \overline {\bf R}_j$. Also, the magnetic superpotential includes cubic couplings $W \propto  q M q + q M_U q$ that will be important for supersymmetry breaking.

\subsection{Standard Model chirality}\label{subsec:chiralSQCD}

The mesons $M_{ij}$ and $(M_U)_{ij}$ defined in (\ref{eq:defM})
have SM quantum numbers given by the antisymmetric part of ${\bf R}_i \times \overline{\bf R}_j$. We wish to identify the composite SM generations
\begin{equation}
T_k\, \in\,{\bf 10}\;,\;\overline F_k \, \in\,\overline {\bf 5}
\end{equation}
with appropriate meson components,
\begin{equation}
T \subset \left(M_{{\bf 10}},\,(M_U)_{{\bf 10}}\right)\;\;,\;\;\overline F \subset \left(M_{\overline{\bf 5}},\,(M_U)_{\overline {\bf 5}}\right)\,.
\end{equation}
In principle it is also possible to use SQCD baryons as composite SM fields, but in the simple examples below this does not lead to realistic models. 

At this stage, composite models can be classified into
\begin{itemize}
\item models where the two composite generations $(T_k, \overline F_k)$, $k=1,2$, arise from the same meson. In the absence of superpotential interactions, there is then an unbroken $U(2)$ flavor symmetry;
\item dimensional hierarchy models, where the lightest first generation arises from the dimension 3 meson $Q U Q$, while the second generation is identified with $Q Q$.
\end{itemize}
We also distinguish between `democratic' models where both ${\bf 10}$ and $\overline {\bf 5}$ are composites, and `ten-centered' models where only the ${\bf 10}$'s are composites. As we discuss below, such models tend to be very minimal and lead to quite realistic flavor textures.

Next, we wish to determine whether it is possible to generate dynamically the SM chirality. This entails calculating the index for the ${\bf 10}$ and $\overline{\bf 5}$ representations
\begin{equation}
\Delta N_{{\bf R}} = N_{\bf R}- N_{\overline{\bf R}}
\end{equation}
where $N_{\bf R}$ is the number of fields $(q, M, M_U)$ in the magnetic theory transforming under the representation ${\bf R}$ of $SU(5)_{SM}$. Even though the SM particles are identified from components of $M$ and $M_U$, the magnetic quarks $q$ are also included in the calculation of the index. The reason for this is that in the single-sector models that we explore, the magnetic gauge group is completely higgsed and the $q$'s couple to some of the components of the mesons, producing vector-like composite messengers. This is explained in more detail in \S \ref{sec:metastable}.

Three different levels of chirality may be distinguished:

a) Models where the Standard Model index is not explained, $\Delta N_{\bf 10}=\Delta N_{\overline{\bf 5}}=0$. It is generated in the UV using spectators. In the absence of spectators the composites are vector-like.

b) Models where the ${\bf 10}+ {\bf \bar 5}$ index is correct, but there are also fermions in other chiral representations. These are better in that, although spectators are still needed, they are in general ``predicted'' by the cancellation of SM anomalies directly in the electric theory.

c) Models where the total index is explained, and it agrees with the ${\bf 10}+ {\bf \bar 5}$ index. The only light chiral fields are then in $\bf{10}$ and $\bf{\bar 5}$ representations, and no spectators are needed besides the usual SM elementary fields ($H$ and the third generation).

These qualifications apply for ten-centered models with simple modifications; the goal in this case is to produce $\Delta N_{\bf 10}=2$ and $\Delta N_{\overline{\bf 5}}=0$ among $(q, M, M_U)$. \S \ref{sec:general} is devoted to the general group theoretic analysis of the SM index. In single-sector models with 2-index representations, we will find a few models of type b) and, surprisingly, just one model of type c) is allowed. 

\subsection{Generating the flavor textures}\label{subsec:flavor}

Let us now explain how the flavor textures are generated~\cite{Franco:2009wf, Craig:2009hf}. At a scale $M_{flavor}>\Lambda$ before the electric theory confines, there is some new dynamics that generates interactions between the elementary Higgs and the mesons, $W \sim M H M$. These operators, being irrelevant in the UV, are suppressed by powers of $M_{flavor}^{-1}$. After confinement they become marginal and give rise to the SM Yukawa couplings, with hierarchies controlled by powers of
\begin{equation}
\epsilon \equiv \frac{\Lambda}{M_{flavor}}\,.
\end{equation}

In models where the first two composite generations (${\bf 10}+{\bf \bar 5}$) come from a single meson $M = QQ$, these new interactions are of the form
\begin{equation}\label{eq:Wdim2}
W_{Yuk} \sim \frac{1}{M_{flavor}^2} (Q Q) H (Q Q) + \frac{1}{M_{ flavor}}(Q Q) H \Psi_3 + \Psi_3 H \Psi_3\,,
\end{equation}
where $\Psi_3$ denotes the elementary third generation. In the IR, after canonically normalizing the meson by $Q Q / \Lambda$, Eq.~(\ref{eq:Wdim2}) gives rise to Yukawa couplings 
\begin{equation}\label{eq:dimtwo}
Y \sim \left( \begin{array}{ccc}
\epsilon^2 & \epsilon^2 & \epsilon \\
\epsilon^2 &\epsilon^2 & \epsilon      \\
\epsilon & \epsilon    & 1 
\end{array}
\right)\,,
\end{equation}
where order one coefficients are being omitted.
For $\epsilon \sim 10^{-1} - 10^{-2}$, these simple Yukawa textures are a good starting point for generating the hierarchies in fermion masses, but more structure is required to obtain fully realistic masses and mixings. In \S \ref{sec:SU2}, we present an alternative mechanism for obtaining realistic Yukawa matrices in models with $U(2)$ flavor symmetry. This will open up new model-building possibilities.

Next, consider dimensional hierarchy models; here the lightest first generation is identified with the dimension 3 meson $M_U = Q U Q$, while the second generation arises from $M= Q Q$. The superpotential at the scale $M_{flavor}$ now reads
\begin{eqnarray}\label{eq:WYukdimH}
W_{Yuk} & \supset & \frac{1}{M_{ flavor}^4} (QU Q) H (QU Q) + \frac{1}{M_{ flavor}^3} (Q Q) H (QU Q) + \nonumber \\
& & \f{1}{M_{ flavor}^2} (Q Q) H (Q Q) + \frac{1}{M_{ flavor}} (Q Q) H \Psi_3 + \Psi_3 H \Psi_3~.
\end{eqnarray}
After confinement, $W_{Yuk}$ gives rise to Yukawa couplings
\be
\label{eq:dimH}
Y \sim \left( \begin{array}{ccc}
\epsilon^4 & \epsilon^3 & \epsilon^2 \\
\epsilon^3 &\epsilon^2 & \epsilon      \\
\epsilon^2 & \epsilon    & 1 
\end{array}
\right)~.
\ee
Realistic flavor textures are obtained for $\epsilon \sim 0.1$~\cite{Craig:2009hf}.

Finally, let us discuss ten-centered models, where the flavor hierarchies come entirely from the ${\bf 10}$'s: the $T_k \in\,{\bf 10}$ corresponding to the first two generations are composites, while $\overline F_k \,\in\,\overline{\bf 5}$ are elementary. As before, in these models the composites $T_1$ and $T_2$ can either be produced by the same dimension two meson, or we can have $T_1 \subset QUQ$, $T_2 \subset QQ$. In the former case, the analog of Eq.~(\ref{eq:Wdim2}) gives
\begin{equation}\label{eq:10Cdimtwo}
Y_u \sim \left( \begin{array}{ccc}
\epsilon^2 & \epsilon^2 & \epsilon \\
\epsilon^2 &\epsilon^2 & \epsilon      \\
\epsilon & \epsilon    & 1 
\end{array}
\right)\;,\;
Y_{d,l} \sim \left( \begin{array}{ccc}
\epsilon & \epsilon & \epsilon \\
\epsilon &\epsilon & \epsilon      \\
1 & 1    & 1 
\end{array}
\right)\,.
\end{equation}
Similarly, for dimensional hierarchy models,
\begin{equation}\label{eq:10CdimH}
Y_u \sim \left( \begin{array}{ccc}
\epsilon^4 & \epsilon^3 & \epsilon^2 \\
\epsilon^3 &\epsilon^2 & \epsilon      \\
\epsilon^2 & \epsilon    & 1 
\end{array}
\right)\;,\;
Y_{d,l} \sim \left( \begin{array}{ccc}
\epsilon^2 & \epsilon^2 & \epsilon^2 \\
\epsilon &\epsilon & \epsilon      \\
1 & 1    & 1 
\end{array}
\right)\,.
\end{equation}
Predictions for masses and mixings are discussed below, after analyzing concrete models.

So far we have assumed that both $H_u$ and $H_d$ are elementary. However, as pointed out in~\cite{SchaferNameki:2010iz}, it is also possible to have a composite $H_d \subset QQ$. This leads to various phenomenologically desirable consequences: the electroweak scale is generated dynamically (and to a smaller extent radiatively), the $\mu/B_\mu$ problem is solved, and the hierarchy between top and bottom/tau masses is naturally explained. While it is not the purpose of this work to discuss in detail the Higgs physics of single-sector models,
we point out that a composite $H_d$ may also lead to attractive ten-centered models, providing an additional suppression for $Y_{d,l}$ in (\ref{eq:10Cdimtwo}), (\ref{eq:10CdimH}).

\section{Chirality and unification: analysis of the SM index}\label{sec:general}

Having explained the gauge dynamics that produces composite generations and the various types of flavor textures that can be obtained, we are now ready to perform a group-theoretic classification of single sector models according to the SM representations that are obtained in the IR and their messenger content. We will compute the index of the complex $SU(5)_{SM}$ representations and use these results to construct models with massless ${\bf \bar 5}$ and/or ${\bf 10}$ composites. This procedure will allow us to find all the examples where perturbative unification may be achieved, and exhibit an interesting connection between SM chirality and unification.

\subsection{Calculation of the index}\label{subsec:index}

Recall that the microscopic theory is SQCD with gauge group $Sp(2N_c)$ and $N_f$ flavors $Q_i$, in the free magnetic range. Absent superpotential interactions, the flavor symmetry group is $SU(2N_f)$, with the electric quarks transforming in the fundamental representation. The SM quantum numbers are explained by weakly gauging a subgroup
$$
SU(5)_{SM} \subset SU(2N_f)\,.
$$
Then the quarks decompose as
\begin{equation}\label{eq:embedQ}
Q \,\sim\, n_1 \times {\bf 1} + n_5 \times {\bf 5}+n_{\bar 5} \times {\bf \bar 5}+n_{10} \times {\bf 10}+n_{\overline{10}} \times {\bf{\overline{ 10}}}\,,
\end{equation}
where the color index is not shown.
For our purposes it is not necessary to consider higher dimensional representations. In general the embedding will be chiral, namely $n_{{\bf r}} \neq n_{{\bf \bar r}}$; SM anomalies are canceled by adding spectators $S_a$ that are singlets under $Sp(2N_c)$. See discussion in \S \ref{subsec:basics}.

In the dual magnetic description, the magnetic quarks decompose into
\begin{equation}\label{eq:embedq}
q \,\sim\, n_1 \times {\bf 1} + n_5 \times {\bf \bar 5}+n_{\bar 5} \times {\bf 5}+n_{10} \times {\bf \overline{10}}+n_{\overline{10}} \times {\bf{ 10}}\,.
\end{equation}
The SM index contributed by the magnetic quarks is then
\begin{eqnarray}\label{eq:indexq}
\Delta N_{10} &\equiv& N_{10}-N_{\overline{10}}= 2 \t N_c(n_{\overline{10}} -n_{10})\nonumber\\
\Delta N_{\bar 5} &\equiv& N_{\bar 5} - N_{5}=2 \t N_c(n_5 -n_{\bar 5})\,,
\end{eqnarray}
where $2 \t N_c$ is the color multiplicity.
Note also that we have defined the index for 5-dimensional representations such that positive index indicates an excess of ${\bf \bar 5}$'s.

The mesons $M_{ij}= Q_iQ_j$ transform in the antisymmetric of $SU(2N_f)$. Taking the antisymmetric part of $QQ$ in Eq.~(\ref{eq:embedQ}) gives:
\begin{eqnarray}\label{eq:indexM}
\Delta N_{50}&=& \frac{n_{\overline{10}}(n_{\overline{10}}-1)} {2}-\frac{n_{10}(n_{10}-1)} {2}\nonumber\\
\Delta N_{45} &=& \frac{n_{\overline{10}}(n_{\overline{10}}+1)} {2}-\frac{n_{10}(n_{10}+1)} {2}+ n_{10} n_{\bar 5} - n_{\overline{10}} n_5\nonumber\\
\Delta N_{40}&=& n_{10} n_5 - n_{\overline{10}} n_{\overline{5}} \nonumber \\
\Delta N_{15}&=&\frac{n_5 (n_5 - 1)}{2} - \frac{n_{\overline{5}} (n_{\overline{5}} - 1)}{2} \nonumber \\
\Delta N_{10} &=&\frac{n_{5}(n_{5}+1)}{2}- \frac{n_{\overline{5}}(n_{\overline{5}}+1)}{2}+ n_1 (n_{10}-n_{\overline{10}})+ n_{\overline{10}} n_{\bar 5}-n_{10} n_5\nonumber\\
\Delta N_{\bar 5} &=&\frac{n_{10}(n_{10}-1)} {2}-\frac{n_{\overline{10}}(n_{\overline{10}}-1)} {2}+n_{\overline{10}}n_5 - n_{10} n_{\bar 5}+ n_1 (n_{\bar 5}- n_5)\,.
\end{eqnarray}
The final expression for the total index associated to the composites $(q, M)$ is thus
\begin{eqnarray}
\Delta N_{50} &=& \frac{n_{\overline{10}}(n_{\overline{10}}-1)} {2}-\frac{n_{10}(n_{10}-1)} {2}\nonumber\\
\Delta N_{45} &=& \frac{n_{\overline{10}}(n_{\overline{10}}+1)} {2}-\frac{n_{10}(n_{10}+1)} {2}+ n_{10} n_{\bar 5} - n_{\overline{10}} n_5\nonumber\\
\Delta N_{40} &=& n_{10} n_5 - n_{\overline{10}} n_{\overline{5}} \nonumber \\
\Delta N_{15} &=&\frac{n_5 (n_5 - 1)}{2} - \frac{n_{\overline{5}} (n_{\overline{5}} - 1)}{2} \nonumber \\
\Delta N_{10} &=&\frac{n_{5}(n_{5}+1)}{2}- \frac{n_{\overline{5}}(n_{\overline{5}}+1)}{2}+ (n_1-2 \t N_c) (n_{10}-n_{\overline{10}})+ n_{\overline{10}} n_{\bar 5}-n_{10} n_5\nonumber
\end{eqnarray}
\begin{eqnarray}\label{eq:Spindex}
\Delta N_{\bar 5} &=&\frac{n_{10}(n_{10}-1)} {2}-\frac{n_{\overline{10}}(n_{\overline{10}}-1)} {2}+n_{\overline{10}}n_5 - n_{10} n_{\bar 5}+ (n_1-2 \t N_c) (n_{\bar 5}- n_5)\,.\;\;\;\;\;
\end{eqnarray}

As we anticipated above, we are computing the total composite index by adding the contributions of $M_{ij}$ and $q_i^\alpha$, although the latter have a magnetic gauge group index $\alpha$ while the former are singlets under $Sp(2 \t N_c)$. We will explain in \S \ref{sec:metastable} that in our models the gauge group is completely higgsed at low energies; the elements of $q$ with nontrivial SM quantum numbers pair with corresponding conjugate fields from $M$ to give vector-like messengers. Accordingly, we introduce the \textit{messenger index} (see e.g.~\cite{Giudice:1998bp})
\begin{equation}\label{eq:Nmess}
N_{mess} \equiv 2 \t N_c ( n_5+n_{\bar 5}+3\, n_{10} + 3\,n_{\overline{10}})\,.
\end{equation}
In what follows, we specialize to the smallest possible magnetic gauge group, namely $\t N_c=1$.

\subsection{Solutions}\label{subsec:sols}

Having determined the index for arbitrary combinations of simple matter, let us now classify the most economical models. For simplicity, we will restrict ourselves to models with $N_f \leq 15$, so that the flavor symmetry is at most $SU(30)$. In order that these models produce successful theories of Standard Model flavor, we will require $ \Delta N_{10} \geq 1$. From the embedding Eq.~(\ref{eq:embedQ}),
\begin{equation}\label{eq:restr}
2N_f = \sum_i\,n_{r_i}\,{\rm dim}({\bf r}_i)
\end{equation}
thus the sum must be even. Finally, in the $Sp$ models of \S \ref{sec:metastable} the supersymmetry breaking vacuum requires 
 $n_1 (n_1 - 1)/2 \ge 1$ or $n_1 \geq 2$, in order not to prematurely break the $SU(5)_{SM}$.

The $Sp$ models with matter $(q,M)$ satisfying the above criteria, ordered by the messenger index $N_{mess}$ of Eq.~(\ref{eq:Nmess}) are:

{\footnotesize
\begin{center}
\begin{tabular}{c|ccccc|cccccc|c}
 & $n_{10}$ & $n_{\overline{10}}$ & $n_5$ & $n_{\bar 5}$ & $n_1$ &$ \Delta N_{50}$ & $\Delta N_{45}$ & $ \Delta N_{40}$ & $\Delta N_{15}$ & $\Delta N_{10}$ & $\Delta N_{\bar 5}$ &  $N_{mess}$\\ \hline

Sp-1 &  0 & 0 & $n_5$ & 0 & $n_1$ & 0 & 0 & 0 & $\frac{n_5 (n_5-1)}{2}$ & $\frac{n_5(n_5+1)}{2}$ & $n_5(2- n_1)$ &  $2n_5$\\

Sp-2 &  1 & 0 & 0 & 0 & $n_1$ & 0 & $-1$ & 0 & 0 & $n_1-2$ & 0 &  6\\

Sp-3 & 0 & 0 & 2 & 1 & $n_1$         & 0 & 0 & 0 & 1 & 2 & $-n_1+2$ &  6 \\

Sp-4 & 1 & 0 & 0 & 1 & $n_1$& 0 &0 & 0 & 0 &$n_1-3$ & $n_1 -3$ & 8 \\ \hline 
Sp-5 & 1 & 0 & 1 & 0 & $n_1$ & 0 & $-1$ & 1 & 0 & $n_1-2$ & $-n_1+2$ & 8 \\ 
Sp-6 & 0 & 0 & 3 & 1 & $n_1$& 0 & 0 & 0 & 3 & 5 & $-2 n_1+4$ &  8 \\

Sp-7 &  1 & 0 & 1 & 1 & $n_1$       & 0 & 0 & 1 & 0 & $n_1-3$ & $-1$ & 10\\
Sp-8 &  0 & 1 & 2 & 0 & $n_1$        & 0 & $-1$ & 0 & 1 & $-n_1+5$ & $-2n_1+6$ &  10 \\
Sp-9 & 0 & 0 & 3 & 2 & $n_1$ & 0 & 0 & 0 & 2 & 3 & $-n_1+2$ & 10\\
Sp-10 & 1 & 0 & 2 & 0 & $n_1$ & 0 & $-1$ & 2 & 1 & $n_1-1$ & $-2 n_1+4$ & 10 \\

Sp-11 & 1 & 0 & 1 & 2 & $n_1$        & 0 & 1 & 1 & $-1$ & $n_1-5$ & $n_1-4$ & 12\\
Sp-12 & 1 & 0 & 2 & 1 & $n_1$ & 0 & 0 & 2 & 1 & $n_1-2$ & $-n_1+1$ &  12  \\
Sp-13 & 2 & 0 & 0& 0 & $n_1$ & $-1$ & $-3$ & 0 & 0 & $2 n_1-4$ & 1 & 12 \\
\end{tabular}
\end{center}
}
The indices in this table are calculated using Eq.~(\ref{eq:Spindex}), which count net number of chiral composites from $q$ and $M$; this classification then applies to single-sector models with $U(2)$ flavor symmetry. Dimensional hierarchy models have additional composites from $M_U$, so that (\ref{eq:indexM}) should be added again. Hoping that it does not lead to confusion, the notation $Sp-k$ from this classification will be used for both $U(2)$ and dimensional hierarchy models. Notice that Eq.~(\ref{eq:restr}) restricts the multiplicities $n_i$. For instance in the class $Sp-1$, $n_1$ and $n_5$ have to be both even or both odd. In class $Sp-2$, $n_1$ must be even, while in $Sp-3$ and $Sp-4$ $n_1$ is odd.

One of our goals is to find models where perturbative unification is possible.
In general, a messenger index of $6$ is the largest that allows perturbative gauge coupling unification if the messengers are between $10^5 - 10^7$ GeV, as is the case for the single-sector models considered here. Therefore we will focus on models $1-3$. Nevertheless, we will also consider model $4$ in detail, as it is the unique model of category (c) (see discussion in \S \ref{subsec:chiralSQCD}). It is useful to explain the salient features of models with small messenger index. After analyzing the supersymmetry breaking mechanism in \S \ref{sec:metastable}, we will study these examples in detail in \S \S \ref{sec:SU2} and \ref{sec:dimh}.
  
Let us begin by considering class $Sp-2$, the unique class of strictly ten-centered models with no extra chiral matter in small representations ($\Delta N_{\bar 5} = 0$). The minimal model of this class that we may construct is a $U(2)$-symmetric model with
$$
n_{10}=1\;,\;n_{\overline{10}} = n_5 = n_{\bar 5} = 0\;,\; n_1=4 \, \\
$$
(the model with $n_1 = 2$ has only ${\bf 10}$'s in messenger fields, but not in the components of the meson in which Standard Model fermions are embedded).  Conveniently, the messenger index is compatible with perturbative gauge coupling unification. Using Eq.(\ref{eq:Spindex}), we find for this model
$$
\Delta N_{10}=2\;,\;\Delta N_{45}=-1\;,\;\Delta N_{50} = \Delta N_{40} = \Delta N_{15} = \Delta N_{\bar 5} =0\,.
$$
This gives a net number of two massless composite in the $\bf{10}$, to be identified with the first two SM generations. These meson components have a $U(2)$ flavor symmetry in the absence of an electric superpotential. Notice that $M$ and $q$ also give two pairs of $\bf{10}+\overline {\bf 10}$ in the minimal case $\t N_c=1$. These become messengers in the supersymmetry breaking model.
Once elementary Standard Model fields have been included, anomaly cancellation requires the addition of spectators to cancel the anomaly contribution of the $\bf{\overline{45}}$. 

We may also use this class to build models with a dimensional hierarchy. In this case, the minimal embedding is identical to that of the $U(2)$ model.  This will lead to extra ${\bf 10}$'s in the pseudomoduli of both meson fields, which must be removed with spectators. However, we will find that this doubling is a desirable feature for models where the doubling of matter fields is required to obtain the correct more-minimal spectrum of MSSM soft masses. 

These are all the examples for ten-centered models with $N_{\bar 5} - N_5 = 0$. But perhaps this is too restrictive and we may have some number of ${\bf 5}$'s lifted by spectators. This is a natural situation in our framework, because the SM generations already give three ${\bf \bar 5}$ spectators. We may then consider ten-centered models in classes $Sp-1$ and $Sp-3$, both of which have chiral ${\bf 5}$'s that must be removed by spectators. 

Consider first class $Sp-1$, which is particularly attractive. The minimal model of this class is 
$$
n_5 = 1 \;,\; n_{10} = n_{\overline{10}} = n_{\bar 5} = 0 \;, \; n_1 = 3\,.
$$
For this model the chiral indices are 
$$
\Delta N_{10} = 1\;,\; \Delta N_{\bar 5} = -1 \; , \; \Delta N_{50} = \Delta N_{45} = \Delta N_{40} = \Delta N_{15} = 0\,.
$$
From the matter content of $q$, two of the ${\bf 5}$'s become messengers in this model, leaving only one ${\bf 5}$ that needs to be lifted by spectators, as can be seen from the index. Of course, since this model has only one ${\bf 10}$, it must necessarily be a dimensional hierarchy model in order to produce the correct number of composite generations. This model is quite compact, with a small messenger index guaranteeing perturbativity at all scales.

What else can we build in class $Sp-1$? There is no model with just two ${\bf 10}$'s. We may build a model with three ${\bf 10}$'s (and $N_{mess} = 4$) via
$$
n_5 = 2 \;,\; n_{10} = n_{\overline{10}} = n_{\bar 5} = 0 \;, \; n_1 = 4
$$
for which the chiral indices are 
$$
\Delta N_{10} = 3\;,\; \Delta N_{\bar 5} = -4 \; , \; \Delta N_{15} = 1 \;,\; \Delta N_{50} = \Delta N_{45} = \Delta N_{40} =  0\,.
$$
The SM anomalies are canceled by adding the following SQCD singlets: one $\overline{\bf 15}$, one $\overline{\bf 10}$ and 6 fields in the $\bf \bar 5$ --two of which are the elementary first and second generation SM fields. Generic superpotential deformations lift the extra unwanted matter at long distance, leaving a confined theory with composites in the $\bf 10$.

We close out our discussion of ten-centered models with those of class $Sp-3$. The simplest such model is 
$$
n_5 = 2 \; , \; n_{\bar 5} = 1 \; , \; n_{10} = n_{\overline{10}} = 0 \;,\; n_1 = 3\,.
$$
The chiral indices from composites in this case are 
$$
\Delta N_{10} =2 \; , \; \Delta N_{\bar 5} = -1 \; , \; \Delta N_{15} = 1 \;,\; \; \Delta N_{50} = \Delta N_{45} = \Delta N_{40} = 0\,.
$$
Here four ${\bf 5}$'s and two ${\bf \bar 5}$'s become messengers. Two ${\bf 24}$'s may be given mass terms, and likewise with a ${\bf 10 + \overline{10}}$ and ${\bf 5 + \overline 5}$ pair. Anomalies are canceled by adding $\overline{\bf 15}$ and ${\bf \bar 5}$ spectators, leaving two chiral ${\bf 10}$'s to become Standard Model states. Such a model can give rise either to a $U(2)$ theory or a dimensional hierarchy with doubled matter, much as in the models of the $Sp-2$ class. This concludes our study of all the simple ten-centered models with minimal spectator content and sufficiently low messenger index.

Now let us analyze models of type c). Here we require all $N_{r}-N_{\bar r}=0$ except for the $\bf{10}$ and/or $\bf 5$. If this can be accomplished, then no spectators are needed and the SM chirality is generated by the gauge dynamics. 
Interestingly, there is a unique $Sp$ example (up to addition of singlets),
\begin{equation}
n_{10}=n_{\bf 5}=1\;,\;n_{\overline{10}}=n_5=0\;,\;n_1 \ge 3\,.
\end{equation}
This corresponds to class $Sp-4$ above, with nonvanishing index
\begin{equation}
N_{10}-N_{\overline{10}}=N_{\bar 5} - N_{5}=n_1-3\,.
\end{equation}
There are no ten-centered models of type c). On the other hand, models in class a) (where $\Delta N_{\bf{10}}=\Delta N_{\bf \bar 5}=0$) were presented in~\cite{Franco:2009wf, Craig:2009hf} so we will not describe them again here.

\vskip 2mm

Let us pause to summarize what we have achieved so far. We have found the $Sp$ gauge theories that give rise to massless meson composites in ${\bf 10}$ and $\bf \bar 5$ SM representations, with smallest messenger number. The masslessness is guaranteed, in the absence of EWSB, by a nonvanishing index computed in the table above. In some of the simpler examples, the number of composite SM families is determined by fields that are \textit{neutral} under $SU(5)_{SM}$. For instance, in classes $Sp-2$ and $Sp-4$ the number of mesons in the ${\bf 10}$ is proportional to the number $n_1$ of electric quarks neutral under the SM gauge group. Similarly, in dimensional hierarchy models the families are associated to mesons $QQ$ and $QUQ$ containing the SM-neutral field $U$ (an antisymmetric of the electric gauge group).\footnote{Class $Sp-3$ is an interesting exception, where a fixed index $\Delta N_{10}$ is obtained from the product of two preons in the ${\bf 5}$.}

Models with SM-chiral electric quarks $Q$ give a small messenger index $N_{mess}$ and make perturbative unification possible. This reveals an interesting connection between SM chirality and unification, the basic reason being that in these models of direct mediation the dynamics that produces composite generations also gives messenger fields. Another consequence of this approach is the existence of spectator fields $S$ that are singlets under the electric $Sp$ gauge group. These are required to make the theory anomaly-free. Once we allow for generic superpotential deformations, these fields will have the desirable effect of lifting extra SM exotics produced by the $Sp$ dynamics. So, let us now study these and other aspects of the low energy theory in detail.

\section{Metastable SUSY breaking in SQCD plus singlets}\label{sec:metastable}

Having explained the flavor hierarchies via compositeness, the next step is to use the same gauge dynamics to break supersymmetry dynamically and calculably. While the original proposals in~\cite{ArkaniHamed:1997fq, Luty:1998vr} were incalculable,~\cite{Franco:2009wf} found that metastable supersymmetry breaking can occur quite naturally in these models. Indeed, ISS showed~\cite{Intriligator:2006dd} that adding small masses to the electric quarks,
\begin{equation}
W_{el}= \tr(m Q \t Q)
\end{equation}
leads, in the free magnetic range, to metastable vacua at the origin $M=0$ of field space. (These formulas refer to an $SU(N_c)$ gauge group; the $Sp$ case will be studied shortly). The macroscopic theory becomes
\begin{equation}
W_{mag}= \tr(mM)+ \frac{1}{\Lambda}\,\tr(q M \t q)
\end{equation}
(where $\Lambda$ is set by the dynamical scale) and supersymmetry is broken by the rank condition.

Single-sector models have one new ingredient, namely additional singlets under the electric gauge group, that can potentially modify the supersymmetry breaking vacua.\footnote{We thank D. Green, A. Katz and Z. Komargodski for pointing out to us that superpotential couplings between electric quarks and singlets in the adjoint of the flavor group produce new metastable vacua with lower energy than the ISS configuration. For further discussion or other applications see~\cite{Green:2010ww}.} We saw in \S \ref{sec:general} that anomaly cancellation in general requires adding spectators $S$ that are neutral under the confining dynamics but have nontrivial $SU(5)$ quantum numbers. In fact, this is phenomenologically attractive because, allowing for generic (gauge invariant) superpotential interactions,
\begin{equation}
W_{el}= \lambda S_{\overline{\bf R}} (Q \t Q)_{\bf R} + \ldots
\end{equation}
can lift unwanted exotics and leave us with just the SM matter fields in the IR.

This naturally leads us to consider SQCD with electric quark masses and couplings to singlets,
\begin{equation}\label{eq:WgenS}
W = (m_i\,\delta_{ij} + \lambda S_{ij}) Q_i \t Q_j
\end{equation}
in the free-magnetic phase.\footnote{The $SU(N_f)$ global symmetry limit was already discussed in~\cite{Essig:2007xk} as a way to generate the quark masses dynamically.} We consider singlets (and masses) that transform nontrivially under the flavor symmetry group, such that the weakly gauged subgroup $SU(5)_{SM}$ of (\ref{eq:embedQ}) is left unbroken. To set some notation, we begin by reviewing the ISS construction (i.e. no extra singlets) for $Sp$ theories with unequal electric masses. Then we study in detail the case of interest Eq.~(\ref{eq:WgenS}); the analysis for $SU$ gauge theories is similar.

\subsection{Metastable vacua in $Sp$ theories with different masses}\label{subsec:review}

As in \S \ref{subsec:basics} we consider an $Sp(2N_c)$ gauge theory with $2N_f$ fundamentals $Q_i^\alpha$, $i=1,\ldots, 2N_f$ in the range $N_c+3 \le N_f < 3(N_c+1)/2$. We also turn on masses for the $N_f$ flavors,
\begin{equation}
W_{el} = \sum_{k=1}^{N_f} \,m_k\,(Q_{2k-1}^\alpha J_{\alpha \beta} Q^\beta_{2k})\,.
\end{equation}
Color indices are contracted with $J_{2N_c}=\mathbf 1_{N_c} \otimes (i \sigma_2)$; subsequent formulas will be sometimes simplified by omitting this contraction.
Quark masses are ordered according to $|m_1| \ge |m_2| \ge \ldots \ge |m_{N_f}|$ and are chosen so that $SU(5)_{SM} \subset SU(2N_f)$ of Eq.~(\ref{eq:embedQ}) is left unbroken; they also have to be parametrically smaller than the dynamical scale, $|m_k| \ll \Lambda$.

The matter content and symmetries of the magnetic theory in the limit $m_k=0$ are
\begin{center}
\begin{tabular}{c|ccc}
&$Sp(2 \t N_c)$&$SU(2N_f)$&$U(1)_R$\\
\hline
$\Phi$& 1 & antisym & 2 \nonumber\\
$q_i^\alpha$& $\Box$ & $\overline \Box$ & 0  
\end{tabular}
\end{center}
where $\t N_c= N_f-N_c-2$ and $\Phi \propto QQ/\Lambda$ and $q$ have canonical kinetic terms. The theory has a superpotential
\begin{equation}
W_{mag}= -h \,\tr (\mu^2 \Phi)+ h \,\tr (\Phi q^T  q)
\end{equation}
where the matrix $\mu^2$ in the linear term is given by
\begin{equation}
\mu^2 = {\rm diag}(\mu_1^2,\ldots,\mu_{N_f}^2) \otimes (i \sigma_2)\;,\;h\mu_i^2 \sim \Lambda m_i\,.
\end{equation}
The diagonal entries are chosen to be real and positive, and ordered so that $\mu_i^2 \ge \mu_j^2$ if $i \ge j$. For simplicity, the cubic coupling $h$ is also taken to be real.

The F-terms
\begin{equation}\label{eq:Fphi}
\frac{\partial W}{\partial \Phi^T} = - h \mu^2 + h (q^T  q)
\end{equation}
cannot all vanish because the second term has a smaller rank than the first. The metastable vacuum is obtained by turning on the maximum number of expectation values to cancel the largest F-terms,
\begin{equation}
\langle q^T q \rangle = {\rm diag} (\mu_1^2,\ldots,\mu_{\t N_c}^2)\otimes (i \sigma_2)\,.
\end{equation}

Fluctuations around the vacuum are parametrized by
\begin{equation}\label{eq:paramPhi}
\Phi= \left(\begin{matrix} Y_{2\tilde N_c \times 2 \tilde N_c} & Z^T_{2\tilde N_c \times 2 (N_f-\t N_c)} \\ - Z_{2(N_f-\t N_c) \times \tilde 2N_c} &X_{2(N_f-\t N_c) \times 2 (N_f-\t N_c)}\end{matrix} \right)\;,\;q^T=\left( \begin{matrix} \chi_{2\tilde N_c \times 2\tilde N_c} \\ \rho_{2(N_f-\t N_c) \times 2\tilde N_c} \end{matrix}\right)\,.
\end{equation}
The tree-level nonzero F-terms and vacuum energy are
\begin{equation}
W_X = -h\, {\rm diag}(\mu^2_{\t N_c+1},\ldots,\mu^2_{N_f}) \otimes (i \sigma_2)\;,\;V_0= 2\sum_{j=\t N_c+1}^{N_f} (h \mu_j^2)^2\,.
\end{equation}
Importantly for what follows, the expectation values $\langle q^T q \rangle$ are set by the largest $\mu_i^2$ and the F-terms are controlled by the smaller ones. The nonzero expectation value $\langle \chi^T \chi\rangle$ higgses completely the magnetic gauge group $Sp(2 \t N_c) \to 1$, and the SM gauge group is a weakly gauged subgroup from
\begin{equation}\label{eq:SU5embedd}
SU(5)_{SM} \subset SU\left( 2(N_f-\t N_c)\right)
\end{equation}
(this group is left unbroken in the limit $\mu_i^2 \to 0$). Notice that $(Z, \rho)$ give $2 \t N_c (\Box + \overline \Box)$ of $SU( 2(N_f-\t N_c))$ and $X$ is an antisymmetric. This justifies our previous procedure for computing the SM index.

The tree-level spectrum is as follows. The field $X$ is a pseudo-modulus; it is flat at tree-level but generically receives quantum corrections and will be lifted. $(Y, \chi)$ are supersymmetric at tree level. On the other hand, $\rho$ couples directly to the pseudo-modulus $X$ which has a nonzero F-term. Also, $\rho$ and $Z$ have a supersymmetric mass $W \supset h \langle \chi \rangle Z \rho$ and have nontrivial SM quantum numbers. Therefore, in the macroscopic theory $(\rho, Z)$ are composite messengers with supersymmetric mass $M= \langle \chi \rangle$ and splittings given by $|W_X|^{1/2}$. More precisely, some of these fields are Nambu-Goldstone modes in the limit $g_{SM} \to 0$; see~\cite{Intriligator:2006dd} for more details.

While this derivation has been for general $\mu_i^2$, for our purposes it suffices to have two different values $\mu_1^2> \mu_2^2$, with the first $\t N_c$ entries equal to $\mu_1^2$ and the last $N_f - \t N_c$ ones equal to $\mu_2^2$. Then the superpotential becomes
\begin{equation}\label{eq:2mus}
W_{mag} = - h\mu_1^2 \,\tr(J_{2\t N_c} Y)- h\mu_2^2 \,\tr(J_{2(N_f-\t N_c)} X) + h \,\tr(\chi Y \chi) + h\,\tr(\rho X \rho)+ 2 h\,\tr(\chi Z \rho)\,,
\end{equation}
with the expectation values simplifying to
\begin{equation}\label{eq:ISSvev}
\langle \chi \rangle = \mu_1\,\mathbf 1_{2 \t N_c}\;,\;W_X = - h \mu_2^2\,\mathbf 1_{2(N_f-N_c)}\;,\;V_0= 2 (N_f- \t N_c) (h^2 \mu_2^4)\,.
\end{equation}
We assume (\ref{eq:2mus}) for the rest of the calculations --the more general case can be approached along similar lines. Here the expectation value of $\chi$ has been fixed by the $Sp$ D-terms.

Integrating out the heavy messengers gives an effective Coleman-Weinberg potential that lifts the pseudo-moduli~\cite{Intriligator:2006dd}
\begin{equation}\label{eq:CW}
V_{CW} = \frac{1}{64 \pi^2}\,{\rm Str}\,\mc M^4 \,\log \frac{\mc M^2}{\Lambda_0^2}= m_{CW} |X|^2+ \mc O(|X|^4)
\end{equation}
where, up to order one numerical factors, the CW mass is
\begin{equation}\label{eq:CWmass}
m_{CW}^2 \approx \frac{h^2}{16 \pi^2}\,\frac{(h \mu_2^2)^2}{\mu_1^2}\,.
\end{equation}
This implies that the MSSM composite sfermions acquire masses of order (\ref{eq:CWmass}).

\subsection{Metastable vacua in SQCD plus singlets}\label{subsec:SSQCD}

Now let us discuss the case relevant for our constructions: an $Sp$ gauge theory with extra singlets $S_{IJ}$ and superpotential
\begin{equation}\label{eq:WSSQCD}
W_{el} = \sum_{k=1}^{N_f} \,m_k\,(Q_{2k-1}Q_{2k})+  \lambda \sum_{I,\,J}\, S_{IJ}(Q_I Q_J)\,.
\end{equation}
Here $(I,J)$ run over an arbitrary subset of the $2N_f$ flavor indices and to avoid a proliferation of indices we have identified all the cubic couplings into a single $\lambda$. The masses and cubic couplings are singlets under the weakly gauged $SU(5)_{SM}$. It is worth exploring first the global limit $g_{SM} \to 0$, which is of more general interest beyond single-sector models. \footnote{SQCD plus singlets has a very rich dynamics in the conformal window; see for instance~\cite{Barnes:2004jj}.} 

The cubic couplings are marginally relevant and their effects are easiest to understand in the magnetic dual, where we have
\begin{equation}\label{eq:SSQCDmag}
W_{mag} = - h\, \tr(\mu^2 \Phi) + h\,\tr( \Phi q^T q) + h \lambda \Lambda\,\tr(S \Phi)\,. 
\end{equation}
The fields $S$ have been grouped into a $2N_f$-rowed antisymmetric matrix, with the understanding that only the elements corresponding to the set $(I,J)$ have nonzero fields. The F-term for $\Phi$ now becomes
\begin{equation}\label{eq:Fmodified}
\frac{\partial W}{ \partial \Phi^T} =-h\, \mu^2 + h (q^T q)+ h \lambda \Lambda\,S\,.
\end{equation}
The rank condition of (\ref{eq:Fphi}) is modified in an important way because the extra singlets make it possible to cancel more than $\t N_c$ F-terms from $h \mu^2$.

Due to the form of the linear terms, it is convenient to split $S$ into `diagonal' and `off-diagonal' pieces,
\begin{equation}\label{eq:Sdef}
S = {\rm diag}(S_1,\ldots,S_{N_f}) \otimes (i \sigma_2) + S'\;,\;\;S'_{k,k+1}=0\,,
\end{equation}
and both contributions can be treated separately. Let us look first at the diagonal terms assuming that there are $n$ nonzero fields $S_a$ and setting $S'=0$. If these elements are all independent, then there are vacua with $\langle S \rangle \sim \mu^2/\lambda \Lambda$ that have lower energy than Eq.~(\ref{eq:ISSvev}). For $n \ge N_f - \t N_c$ these vacua are supersymmetric --here we focus on the supersymmetry breaking case $n < N_f - \t N_c$. 

If $S$ has $n_1$ elements in the upper $2\t N_c$ block (the analog of $Y$ in Eq.~(\ref{eq:paramPhi})) and $n_2$ elements in the lower $2(N_f - \t N_c)$ block, with $n_1 < \t N_c$ and $n_1+n_2 < N_f- \t N_c$, the new supersymmetry breaking vacuum has
\begin{eqnarray}
\langle S_1 \rangle & = & \ldots = \langle S_{n_1}\rangle= \frac{\mu_1^2}{\lambda \Lambda}\;,\;
\langle S_{n_1+1} \rangle = \ldots =\langle S_{n_1+n_2} \rangle= \frac{\mu_2^2}{\lambda \Lambda}\nonumber\\
\langle \chi^T \chi \rangle & = & \mu_1^2\,J_{2(\t N_c-n_1)}\;,\;\langle \rho^T \rho \rangle = \mu_2^2\,J_{2(N_f-\t N_c-n_1)}\,.
\end{eqnarray}
This gives an energy
\begin{equation}
V_0 = 2 (N_f-\t N_c-n_1-n_2) ,h^2 \mu_2^4
\end{equation}
lower than the ISS state before. These extrema correspond to canceling $n_1+n_2$ masses of electric quarks. The two-loop instability of SQCD with massive and massless quarks~\cite{Franco:2006es, Giveon:2008wp} is absent here because the mesons with vanishing linear term have a large $\mc O(\lambda \Lambda)$ mass.

Still in the situation with $S'=0$, one interesting case (that will appear below) is when there are restrictions on the components of $S$. For instance, we can impose the `traceless' condition on the $n$ nonvanishing diagonal elements in (\ref{eq:Sdef}),
\begin{equation}\label{eq:Sconstr}
\tr(J_{2n} S) = 0\;\Rightarrow\;\sum S_j=0\,.
\end{equation}
This arises quite naturally in $Sp$ examples. Let us further assume that the nonvanishing $S_a$ sit in the lower $2(N_f- \t N_c)$ block, parallel to $X$ in Eq.~(\ref{eq:paramPhi}). This leads to a rich set of vacua because now turning on $\langle S \rangle $ also requires nonzero $\langle \rho^T \rho \rangle$ in order to decrease the extra energy contribution from the constraint. However, by the rank condition, some of the components in $\langle \chi^T \chi \rangle$ have to be turned off. As a result, the vacuum energy depends on both $\mu_1$ and $\mu_2$, as well as $n$.

In more detail, the new metastable configuration has
\begin{eqnarray}\label{eq:newVEV}
\langle \chi^T \chi \rangle &=& \mu_1^2\,J_{2 (\t N_c-1)}\;,\;\langle \rho^T \rho \rangle = \mu_2^2\,{\rm diag} (0,\ldots,0,n) \otimes (i \sigma_2)\nonumber\\
\langle S \rangle &=& \frac{\mu_2^2}{\lambda \Lambda}\,{\rm diag}(1,\ldots,1,-(n-1)) \otimes(i \sigma_2)\,.
\end{eqnarray}
There are non-vanishing F-term from both $Y$ and $X$, giving
\begin{equation}\label{eq:V0new}
V_0 = 2 \,h^2 \mu_1^4+2(N_f-\t N_c-n)\, h^2 \mu_2^4\,.
\end{equation}
Recalling that $\mu_1^2 > \mu_2^2$, the new vacua have lower energy than the ISS configuration (\ref{eq:ISSvev}) for
\begin{equation}
\mu_2^4 < \mu_1^4 < n\,\mu_2^4
\end{equation}
while the metastable vacuum with vanishing $S$ is preferred for $\mu_1^4 > n\,\mu_2^4$.

In the limiting case $\mu_1/\mu_2\to 1$, the new configuration has lower energy. It would be interesting to compute the rate of decay from (\ref{eq:ISSvev}) to (\ref{eq:newVEV}) and understand whether the life-time can be made realistically long. This requires a numerical evaluation because there is no small parameter controlling the bounce action.

Finally, let us include the off-diagonal contributions $S'$. If there is only one singlet
$$
S'_{ab}\;,\; (a,b) \neq (k,k+1)\;,
$$
then the critical point of the F- plus D-term potential is at $\langle S' \rangle =0$ and there are no new vacua around the origin $\Phi=0$. New metastable vacua appear when there are at least two singlets,
\begin{equation}
W_{mag} \supset h \Lambda \left(\lambda S'_{ab} \Phi_{ba} + \t \lambda S'_{a+1,b+1} \Phi_{b+1,a+1} \right)\;,\; (a,b) \neq (k,k+1)\,.
\end{equation}
Indeed, now we can combine the reduced rank matrix $q^T q$ with the singlets to cancel  more than $\t N_c$ F-terms. Assuming for instance that $a,b > \t N_c$ (so that the singlets couple to $X$) there is a vacuum with
\begin{equation}
 \chi^T \chi  = \mu_1^2 J_{2(\t N_c-1)}\;,\;(\rho^T \rho)_{a,a+1}=-(\rho^T \rho)_{b,b+1}= \mu_2^2
\end{equation}
and
\begin{equation}\label{eq:chiralvacua}
S'_{ab} = - \frac{\mu_2^2}{\lambda \Lambda}\;,\;S'_{a+1,b+1} =  \frac{\mu_2^2}{\t \lambda \Lambda}\,.
\end{equation}
Imposing the $Sp$ D-terms further restricts $|\rho_a|=|\rho_{a+1}|=\mu_2$ and similarly for the other magnetic quarks. 

In this configuration, two F-terms from $X$ are cancelled instead of one F-term for $Y$, resulting in a vacuum energy
\begin{equation}
V_0 = 2 \,h^2 \mu_1^4+ 2 (N_f-\t N_c-2) \,h^2 \mu_2^4\,.
\end{equation}
The new vacuum is energetically preferred over the ISS one for
\begin{equation}
\mu_1^4< 2\, \mu_2^4\,.
\end{equation}
This result generalizes readily to $n$ singlets with the above couplings. We will see shortly that these vacua associated to off-diagonal spectators disappear after weakly gauging $SU(5)_{SM}$.

\subsection{Single-sector supersymmetry breaking}\label{subsec:ssDSB}

Returning to single-sector models, we now turn on $g_{SM} \neq 0$.
This introduces additional D-term contributions that have to be minimized together with the F-terms and magnetic D-terms. Unlike vacua of \S \ref{subsec:review} (which are along SM D-flat directions), the new configurations of \S \ref{subsec:SSQCD} involve expectation values for $S$ and $\rho$ that are charged under $SU(5)$.

Examples with nonzero diagonal elements $S_a$ in Eq.~(\ref{eq:Sdef}) arise,
for instance, from spectators in the adjoint of $SU(5)_{SM}$. Though not required by anomaly cancellation, these fields may be useful if there is extra matter in the ${\bf 24}$ of $QQ$ that needs to be lifted. The configuration~(\ref{eq:newVEV}) extremizes the SM D-terms, so it gives a minimum in the theory with nonvanishing $g_{SM}$. Therefore, in the case of equal masses (or more generally for $\mu_1^4< n \,\mu_2^4 $ with $n=5$ here) there are new states with lower vacuum energy than the ISS configuration. 

In particular, this means that in some of the explicit models in~\cite{Franco:2009wf, Craig:2009hf} (that included spectators in the adjoint) there are additional vacua with lower energy. Decay rates to these new configurations are not parametrically suppressed, and a numerical calculation of the life-time is needed. It is certainly possible that the ISS-type vacua can still be realistically long-lived, and it would be interesting to understand this in more detail. In any case, a simple fix is to take different electric quark masses, as explained before. Indeed, the ISS vacuum is energetically preferred already for order one differences $\mu_1^2 \ge \sqrt{5} \,\mu_2^2$, and the tunneling instability to the new vacua is then absent.

On the other hand, spectators in chiral representations of $SU(5)$, required for anomaly cancellation, correspond to $S'$ in Eq.~(\ref{eq:Sdef}). For instance, in class $Sp-1$ there is a $\bf \bar 5$ spectator, with
$$
W_{el} \supset \,\lambda \,S_{\bf \bar 5} (QQ)_{\bf 5}\,.
$$
This case is qualitatively different from the previous one because the SM D-terms can no longer be minimized at $\langle S \rangle \neq 0$. (For this conclusion to hold it is also important to impose the $Sp$ D-terms). Hence for chiral spectators there are no new vacua (\ref{eq:chiralvacua}) once $g_{SM}$ is nonzero.\footnote{One could try to cancel the nonzero D-terms by using additional elementary fields that may be present in particular models (e.g. the elementary $\overline F_a$ in ten-centered models). However, minimizing the contributions from F-terms and $Sp(2 \t N_c) \times SU(5)_{SM}$ D-terms shows that there are no such vacua near the origin $\Phi=0$. We thank D. Green for very interesting discussions on these possibilities.}

For model-building purposes, we can then just focus on the ISS vacuum --the additional states of \S \ref{subsec:SSQCD} are either absent in chiral models or can be easily made energetically disfavored.\footnote{Note that for the case of spectators in the adjoint, the vacua of Eq.~(\ref{eq:newVEV}) completely break the SM gauge group. They could lead to interesting phenomenology in models where for instance $SU(5)_{SM}$ is embedded in the $SU(2 \t N_c)$ flavor subgroup. A possible application was recently discussed in~\cite{Green:2010ww}.} The chiral exotics are rendered massive via couplings to spectators, acquiring a mass $\lambda \Lambda$ near the compositeness scale. On the other hand, the SM composite generations (identified with some of the elements of $X$) couple directly to the composite messengers through (\ref{eq:2mus}) and acquire positive one-loop squared masses
\begin{equation}
m_{CW}^2 \sim \frac{h^2}{16 \pi^2}\,\frac{h^2 \mu_2^4}{\mu_1^2}\,.
\end{equation}
(The pattern of soft masses in dimensional hierarchy models is slightly more involved and has been analyzed in~\cite{Craig:2009hf}).

Elementary sfermions receive only two-loop gauge mediated masses,
\begin{equation}
m_{GM}^2 \sim \left( \frac{g_{SM}^2}{16 \pi^2}\right)^2\,\frac{h^2 \mu_2^4}{\mu_1^2}\,.
\end{equation}
Gaugino masses are obtained by deforming the electric theory with a quartic operator $(QQ)^2$, which leads in the IR to
\begin{equation}
W_{mag} \supset \frac{h^2}{2} \muphi\, \tr \,\Phi^2\,.
\end{equation}
This explicitly breaks the $U(1)_R$ symmetry and produces a (loop-enhanced) expectation value for $X$ leading to gaugino masses~\cite{Essig:2008kz}:
\begin{equation}
\langle h X \rangle \sim 16 \pi^2 \,\muphi\,\frac{\mu_1^2}{\mu_2^2}\;\Rightarrow\; m_\lambda \sim g_{SM}^2 \muphi\, \frac{\mu_2^4}{\mu_1^4}\,.
\end{equation}

Models where both $T_a$ and $\overline F_a$ are composites present a ``more minimal'' type spectrum~\cite{Cohen:1996vb}. In ten-centered models only the composite $T_k$ acquire heavy CW masses, while all the $\overline F_a$ (as well as the complete third generation) sfermions and gauginos are much lighter. For
\begin{equation}
\frac{h \mu_1^2}{\mu_2} \sim 100 - 200\;{\rm TeV}\;\;,\;\;\muphi \sim\; 1\;{\rm TeV}
\end{equation}
composite masses are of order $10 - 20$ TeV, while the elementary fields are at the TeV scale. For this choice of parameters the LSP gravitino has a mass
\begin{equation}
m_{3/2} \sim \frac{F}{\sqrt 3\,M_{Pl}} \sim \,1 - 10\;{\rm eV}\,,
\end{equation}
satisfying cosmological constraints. We refer the reader to~\cite{Essig:2008kz, Craig:2009hf, SchaferNameki:2010iz} for a detailed analysis of supersymmetry breaking and soft spectra.

\section{Models with $U(2)$ flavor symmetry}\label{sec:SU2}

We have now gathered all the necessary tools to construct realistic chiral single sector models. This section presents the analysis of models where both composite generations are obtained from elements of the same meson. Dimensional hierarchy models are studied in \S \ref{sec:dimh}.

Models with composite generations from dimension 2 mesons are attractive for various reasons. The supersymmetry breaking sector is quite simple and the generation of realistic soft masses, straightforward. The number of SM exotics at high scales is comparatively small, and perturbative unification ensues. Moreover, these models give a very elegant solution to the flavor problem, combining decoupling and universality between the composite generations, enforced by the $U(2)$ flavor symmetry. (Contributions to flavor-changing neutral currents are analyzed in the appendix). However, these constructions do not give fully realistic fermion textures. While the larger hierarchies are naturally explained --i.e. why the third generation fermions are much heavier than the ones from the first two generations-- still Eqs.~(\ref{eq:dimtwo}) or (\ref{eq:10Cdimtwo}) do not predict the correct spectrum for lighter fermions.

We will now argue that the situation is remedied by requiring that the physics at the scale $M_{flavor}$ (responsible for generating the higher-dimensional interactions between the Higgs and mesons) also respects a $U(2)$ symmetry that acts on the first two composite generations $(\Psi_1, \Psi_2)$. Flavor-invariant interactions are constructed by adding a set of scalars $\phi$ that transform under appropriate representations of $U(2)$, and are singlets under the confining dynamics. The expectation values of these ``flavons'' spontaneously break the flavor symmetry and will be used to obtain realistic textures.

The idea of using these flavons in combination with various patterns of flavor symmetry breaking is well-known.\footnote{In particular, we refer the reader to~\cite{flavorU2} for an analysis of models with $U(2)$ flavor symmetry.} Our point here is that single-sector models where a dimension 2 meson gives rise to the SM composites can fruitfully combine the generation of textures via compositeness with the requirement of a $U(2)$ symmetry on the Yukawa interactions. It is also necessary to point out that in ten-centered models it is natural to have a $U(2) \times U(2)$ acting on $(T_1, T_2)$ and $(\overline F_1, \overline F_2)$ separately. Here we focus on the particular case of a single $U(2)$ with the aim of proving in a simple setup that this class of single-sector models can lead to realistic textures. It would be interesting to study in generality the new types of flavor patterns that are possible. In particular, while for simplicity the flavons will be chosen as singlets of $SU(5)_{SM}$, it would be nice if some of the spectator fields could also play a useful role in flavor interactions.

A possible realistic scenario requires one flavon $\phi^a$ in the fundamental representation and a symmetric tensor $\phi^{ab}$. Their expectation values are chosen to be of the form
\begin{equation}
\langle \phi^a \rangle = \delta^{a2}\,\Lambda v_2\;\;,\;\;\langle \phi^{ab} \rangle = (\delta^{a1} \delta^{b2}+ \delta^{a2} \delta^{b1})\,\Lambda\,v_1
\end{equation}
where the dynamical scale $\Lambda$ has been included for later convenience and $v_1$ and $v_2$ are dimensionless.  We should stress that here we do not explain how these expectation values are obtained, although it would be nice to find a mechanism that relates them to the natural scales $\Lambda$ or $M_{flavor}$. The largest flavor hierarchies are already generated via compositeness, suggesting that the flavon dynamics could be quite simple.

In `democratic' models where each full generation $\Psi_a=T_a + \overline F_a$ is composite, combining (\ref{eq:dimtwo}) with $U(2)$ invariance yields
\begin{equation}\label{eq:WyukU2}
W_{Yuk}=  \epsilon^2 \frac{\phi^a \phi^b}{M_{flavor}^2}\, \Psi_a H \Psi_b + \epsilon^2 \frac{\phi^{ab}}{M_{flavor}} \Psi_a H \Psi_b + \epsilon \frac{\phi^a}{M_{flavor}}\,\Psi_3 H \Psi_a+ \Psi_3 H \Psi_3\,.
\end{equation}
Eq.~(\ref{eq:WyukU2}) gives rise to the following Yukawa textures (up to order one factors)
\begin{equation}
Y \sim \left( \begin{array}{ccc}
0 & v_1\epsilon^3  & 0 \\
v_1\epsilon^3   & v_2^2 \epsilon^4 & v_2  \epsilon^2     \\
0 & v_2 \epsilon^2      & 1 
\end{array}
\right)~.
\end{equation}
In particular, fermion masses obey the relations
\begin{equation}
\frac{m_c}{m_t} \sim v_2^4\,\epsilon^4\;\;,\;\;\frac{m_u}{m_c} \sim \frac{v_1^2}{v_2^4 \epsilon^2}\, ,
\end{equation}
while similar relations hold for down-type quarks. A realistic spectrum can then be obtained for
\begin{equation}
v_1 \sim \epsilon \sim 10^{-1}\;,\;v_2 \gtrsim \mc O(1)\,.
\end{equation}

For ten-centered models, on the other hand, we obtain
\begin{eqnarray}
W_{Yuk} &=&  \epsilon \frac{\phi^a \phi^b}{M_{flavor}^2}\,T_a ( \epsilon H_u T_b + H_d \overline F_b) +\epsilon \frac{\phi^{ab}}{M_{flavor}} T_a (\epsilon H_u T_b + H_d \overline F_b) \nonumber\\
&+&  \frac{\phi^a}{M_{flavor}}\,(\epsilon T_a H_u \overline F_3+T_3 H_d \overline F_a)+ T_3 H_u T_3 + T_3 H_d \overline F_3\,.
\end{eqnarray}
In this case the up-type Yukawas are unaltered from the previous case, while the down-type Yukawas read
\begin{equation}
Y_{d} \sim \left( \begin{array}{ccc}
0 & v_1\epsilon^2 & 0 \\
v_1\epsilon^2  & {v_2}^2 \epsilon^3  & v_2 \epsilon^2      \\
0 &  v_2 \epsilon    & 1 
\end{array}
\right)~.
\end{equation} 
Numerical examples for fermion masses are reserved for Appendix \ref{sec:FCNC}.

\subsection{Ten-centered models}\label{subsec:10c}

Thus far we have explored the general features of supersymmetry breaking and the flavor hierarchy in single-sector models based on $Sp(N)$ SQCD. Returning now to the chiral models classified in \S~\ref{subsec:sols}, we may build explicit theories of dynamical supersymmetry breaking and chiral flavor. Let us begin with perhaps the simplest of all explicit models: ten-centered theories with $U(2)$ flavor symmetry.

\subsubsection*{A model of $Sp-1$}

Consider first an $Sp(2 N_c = 8)$  gauge theory with $N_f = 7$ flavors of fundamentals. This corresponds to class $Sp-1$ in the notation of \S~\ref{subsec:sols}. The flavor symmetry of the theory is $SU(14)$  with $Sp(14)$ left as the diagonal global symmetry. The dual theory -- as is the case for all models in this section -- is an IR free $Sp(2 \tilde N_c)$ gauge theory with $\tilde N_c = N_f - N_c - 2 = 1$ and $2N_f$ magnetic quarks transforming as conjugates of the electric quarks, plus the gauge singlet meson.

The embedding of $SU(5)$ in the flavor symmetry of the UV theory is
\begin{center}
\begin{tabular}{c|cc}
&$Sp(8)$&$SU(5)_{SM}$\\
\hline
$Q_i^\alpha$& $\Box$ & $({\bf 5} + {\bf 5} + {\bf 1} + {\bf 1})+ {\bf 1} +{\bf 1}$  \nonumber\\
$S_a$& $\bf 1$ & ${\bf \overline{15}} + {\bf \overline{10}} + 4 \times {\bf \overline 5} $
\end{tabular}
\end{center}
where the parenthesis denotes the $Sp(12)$ subgroup of the flavor symmetry that remains unbroken in the nonsupersymmetric vacuum. To simplify the exposition, the anomaly-free elementary SM matter content is omitted in the list of spectator fields. The product gauge group theory is furthermore perturbed by a renormalizable superpotential
\begin{equation}\label{eq:Sp-Wpert}
W = m_1 (Q_1Q_2)+ m_2 \sum_{k=2}^6 (Q_{2k-1} Q_{2k})+ \lambda \sum_a S_a (Q Q)_{\bar a}
\end{equation}
where (as explained before) it is enough to consider two different electric masses.

The structure of the magnetic dual is
\begin{center}
\begin{tabular}{c|cc}
&$Sp(2)$&$SU(5)_{SM}$\\
\hline
$M_{ij}$& $\bf 1$ & $8 \times {\bf 5} + 3 \times {\bf 10} + {\bf 15} + 6 \times {\bf 1}$  \nonumber\\
$q_i^\alpha$& $\Box$ & $(\bar{\bf 5} + \bar{\bf 5}  + {\bf 1} + {\bf 1})+ {\bf 1} +{\bf 1}$  \nonumber\\
$S_a$& $\bf 1$ & ${\bf \overline{15}} + {\bf \overline{10}} + 4 \times {\bf \overline 5}$
\end{tabular}
\end{center}
The resulting composite messengers comprise
\begin{equation}
(\rho \oplus Z) \sim 2 \times (2 \times {\bf 5} + 2 \times {\bf \bar 5})
\end{equation}
and the lower block of the meson transforms as a $12 \times 12$ antisymmetric tensor, decomposing under $SU(5)$ as
\begin{equation}
X \sim 2 \times {\bf{10}} + \left[{\bf 10} + {\bf 15} + 4 \times {\bf 5} + {\bf 1} \right]\,.
\end{equation}

The fields $(S_a, X_{\bar a})$ acquire masses of order $\lambda \Lambda$ and decouple from the low energy theory. The electric mass terms lead to the metastable vacuum of \S \ref{subsec:review}. The only massless composites in the IR are then the first two ${\bf 10}$'s in $X$, giving the required SM matter fields. Given this matter content, the messenger index for this theory is $N_{mess} = 4$, more than compatible with perturbative gauge coupling unification (keeping in mind that there are significant additional contributions at the scale $\lambda \Lambda$).


\subsubsection*{A model of $Sp-2$}

Consider now an $Sp(2 N_c = 8)$  gauge theory with $N_f = 7$ flavors of fundamentals. In the limit of vanishing superpotential, the flavor symmetry of the theory is $SU(14)$ with $Sp(14)$ left as the diagonal global symmetry.  The embedding in the UV theory is 
\begin{center}
\begin{tabular}{c|cc}
&$Sp(8)$&$SU(5)_{SM}$\\
\hline
$Q_i^\alpha$& $\Box$ & $({\bf 10} +{\bf 1} + {\bf 1})+ {\bf 1} +{\bf 1}$  \nonumber\\
$S_a$& $\bf 1$ & ${\bf 45}$
\end{tabular}
\end{center}
where again the SM elementary fields are not shown, and a perturbation analogous to Eq.~(\ref{eq:Sp-Wpert}) is turned on.

The structure of the magnetic dual is
\begin{center}
\begin{tabular}{c|cc}
&$Sp(2)$&$SU(5)_{SM}$\\
\hline
&&\\[-12pt]
$M_{ij}$& $\bf 1$ & $4 \times {\bf 10} + {\bf \overline{45}} + 6 \times {\bf 1}$  \nonumber\\
$q_i^\alpha$& $\Box$ & $(\overline{\bf 10} +{\bf 1} + {\bf 1})+ {\bf 1} +{\bf 1}$  \nonumber\\
$S_a$& $\bf 1$ & ${\bf 45}$
\end{tabular}
\end{center}
For the model considered here, we have composite messengers
\begin{equation}
(\rho \oplus Z) \sim 2 \times ({\bf 10} + {\bf {\overline {10}}}+ 4 \times {\bf 1})
\end{equation}
and the lower block of the meson transforms as a $12 \times 12$ antisymmetric tensor, decomposing under $SU(5)$ as
\begin{equation}
X \sim 2 \times {\bf{10}} + \left[{\bf{\overline{45}}} + {\bf 1} \right]\,.
\end{equation}
Given this matter content, the messenger index for this theory is $N_{mess} = 6$, just compatible with perturbative gauge coupling unification.

To this set of fields we must add the usual complement of elementary Standard Model fields: two ${\bf \bar{5}}$ for the first two generations, as well as one ${\bf \bar 5 + 10}$ pair for the elementary third generation. Given this field content, we must also add one ${\bf 45}$ in order for the theory to be anomaly-free. Conveniently, this pairs with the ${\bf \overline{45}}$ contained in $X$ to obtain a mass at the duality scale, leaving no superfluous fields charged under $SU(5)$ at low energies. In this case it is amusing to note that the massive ${\bf 45}$ may be used to generate a Georgi-Jarlskog texture \cite{Georgi:1979df} for the mass matrix of the Standard Model fermions. 

\subsubsection*{A model of $Sp-3$}

We may also build a somewhat less attractive ten-centered meson model using the embedding $Sp-3$. Consider an $Sp(2 N_c = 12)$  gauge theory with $N_f = 9$ flavors of fundamentals. The flavor symmetry of the theory is $SU(18)$ embedding with $Sp(18)$ left as the diagonal global symmetry. The embedding of $SU(5)$ in the flavor symmetry of the UV theory is

\begin{center}
\begin{tabular}{c|cc}
&$Sp(12)$&$SU(5)_{SM}$\\
\hline
&&\\[-12pt]
$Q_i^\alpha$& $\Box$ & $({\bf 5} + {\bf 5} + {\bf \bar 5} + {\bf 1})+ {\bf 1} +{\bf 1}$  \nonumber\\
$S_a$& $\bf 1$ & $\overline{\bf 15} + \overline{\bf 5}$
\end{tabular}
\end{center}

The structure of the magnetic dual is
\begin{center}
\begin{tabular}{c|cc}
&$Sp(2)$&$SU(5)_{SM}$\\
\hline
&&\\[-12pt]
$M_{ij}$& $\bf 1$ & $6 \times {\bf 5} + 3 \times {\bf \bar 5} + 3 \times {\bf 10} +\overline{\bf 10} + {\bf 15} + 2 \times {\bf 24}  + 5 \times {\bf 1}$  \nonumber\\
$q_i^\alpha$& $\Box$ & $({\bf 5} + {\bf \bar 5} + {\bf \bar 5} + {\bf 1})+ {\bf 1} +{\bf 1}$   \nonumber\\
$S_a$& $\bf 1$ & $\overline{\bf 15} + \overline{\bf 5}$
\end{tabular}
\end{center}

The composite messengers for this model comprise
\begin{equation}
(\rho \oplus Z) \sim 2 \times (3 \times {\bf 5} + 3 \times {\bf \bar 5} + 2 \times {\bf 1})
\end{equation}
and the lower block of the meson transforms as a $16 \times 16$ antisymmetric tensor, decomposing under $SU(5)$ as
\begin{equation}
X \sim 2 \times {\bf{10}} + \left[({\bf 10 + \overline{10}}) + ({\bf 5 + \bar 5}) + {\bf 5} + {\bf 15} + 2 \times {\bf 24}  + 2 \times {\bf 1} \right]\,.
\end{equation}
Given this matter content, the messenger index for this theory is $N_{mess} = 6$, just compatible with perturbative gauge coupling unification.

\subsection{Model without spectators}\label{subsec:noS}

Finally among the $U(2)$-symmetric theories, let us turn to the unique model with chiral SM fields and no need for spectators ($Sp-4$ in the classification scheme of \S~\ref{subsec:sols}). In contrast to the ten-centered models considered above, these spectator-free models automatically contain both ${\bf 10}$ and ${\bf \bar 5}$ representations.

The most minimal such model has a magnetic gauge group with $\t N_c=1$ and two composite SM generations; this corresponds to
\begin{equation}\label{eq:choice}
N_f=10\;,\;N_c=7\;,\;n_1=5\,.
\end{equation}
The UV theory is 
\begin{center}
\begin{tabular}{c|cc}
&$Sp(14)$&$SU(5)_{SM}$\\
\hline
&&\\[-12pt]
$Q_i^\alpha$& $\Box$ & $({\bf 10 + \bar 5} + {\bf 1 + 1 + 1})+ {\bf 1} +{\bf 1}$  \nonumber\\
$S_a$& $\bf 1$ & $-$
\end{tabular}
\end{center}
with magnetic dual
\begin{center}
\begin{tabular}{c|cc}
&$Sp(2)$&$SU(5)_{SM}$\\
\hline
&&\\[-12pt]
$M_{ij}$& $\bf 1$ & $5 \times {\bf{10}} + 5 \times {\bf {\bar 5}} + {\bf 45}+{\bf{\overline{45}}}+ {\bf{\overline{10}}}+ {\bf 5}+ 10 \times {\bf 1}$  \nonumber\\
$q_i^\alpha$& $\Box$ & $({\bf \overline{10} + 5} + {\bf 1 + 1 + 1})+ {\bf 1} +{\bf 1}$  \nonumber\\
$S_a$& $\bf 1$ & $-$
\end{tabular}
\end{center}

The composite messengers in this theory consist of 
\begin{equation}
(\rho \oplus Z) \sim 2 \times ({\bf 10} + {\bf {\overline {10}}}+ {\bf 5} + {\bf{\bar 5}}+ 3 \times {\bf 1})
\end{equation}
and the lower block of the meson transforms as
\begin{equation}
X \sim 2 \times ( {\bf{10}} + {\bf {\bar 5}}) + \left[{\bf 45}+{\bf{\overline{45}}}+{\bf{10}} + {\bf {\bar 5}}+ {\bf{\overline{10}}}+ {\bf 5}+ 3 \times {\bf 1} \right]\,.
\end{equation}
Excluding the singlet necessary for supersymmetry breaking, the extra representations inside the brackets are vector-like and are made massive by deforming
$$
\Delta W \propto \tr\,X^2
$$
This deformation gives masses of order $\epsilon \Lambda$ to the extra vector-like matter. As discussed in \S \ref{sec:metastable}, the metastable vacuum is stabilized by allowing for different electric quark masses.

Unfortunately, the messenger index for this theory is $N_{mess} = 8$, rendering it incompatible with perturbative gauge coupling unification; SM gauge couplings hit a Landau pole around $10^{12}$ GeV in this theory, assuming messengers around 250 TeV. Given that in our analysis this is the unique model without spectators, it would nice to improve this situation, perhaps along the lines of~\cite{Abel:2008tx}.

\section{Models with dimensional hierarchy}\label{sec:dimh}

Next, we consider models where a realistic flavor structure is generated directly in the UV theory without the addition of flavor spurions. To accomplish this, we consider a variation of the $Sp(N)$ model, adding an antisymmetric tensor~\cite{Intriligator:1995ff}; this leads to the $Sp(N)$ analog of the $SU(N)$ dimensional hierarchy models constructed in \cite{Craig:2009hf}.

The electric theory is $Sp(2N_c)$ ($\subset SU(2N_c)$), with $N_f$ flavors ($Q_i$, $i=1,\ldots, 2N_f$) and a field $U$ in the ``traceless'' antisymmetric ($N_c(2N_c-1)-1$) of the gauge group. The antisymmetric field has a superpotential
\begin{equation}
W=\frac{g_U}{3} \Tr (J_{2N_c}U)^3 + \frac{m_U}{2} \Tr (J_{2N_c}U)^2 + \lambda \Tr (J_{2N_c}U)\,.
\end{equation}
which restricts the mesons to
\begin{equation}
M = Q Q\;,\;M_U = QUQ\,.
\end{equation}
Both are in the antisymmetric of the flavor group $SU(2N_f)$. The Lagrange multiplier $\lambda$ enforces the ``traceless'' condition, setting to zero the Sp singlet $(J_{2N_c}U)=0$.

This corresponds to $k=2$ in the superpotential $W= \Tr U^{k+1}$ of~\cite{Intriligator:1995ff}; we include a superpotential mass term for phenomenological reasons. The dual is $Sp(2 \t N_c)$ with
$$
\t N_c \equiv k(N_f-2) -N_c=2(N_f-2)-N_c\,.
$$
As with the $U(2)$-symmetric models considered above, we will restrict to the minimal case $\t N_c=1$, for which $Sp(2)=SU(2)$. 

The dual has a ``magnetic'' traceless antisymmetric $\t U$, $2N_f$ fundamentals $q$, and canonically normalized singlets $\Phi$ and $\Phi_U$ corresponding to the above mesons. For $\t N_c=1$, the theory does not contain the field $\t U$ -- the antisymmetric is just a singlet, and this vanishes by the traceless condition. This is the analog of s-confining SQCD. Then the magnetic superpotential simplifies to
\begin{equation}
W_{mag} = h_1 \tr (q J_{2 \t N_c} q \Phi) + h_2 \tr (q J_{2 \t N_c} q \Phi_U)
\end{equation}
where $h_1/h_2 \propto m_U/(g_U \Lambda)$, $\Lambda$ being the dynamical scale. We will break supersymmetry by the addition of a deformation whose IR form is
\begin{equation}
W \supset - h_2 \mu^2 \tr (\Phi_U J_{2 N_f})
\end{equation}
The effect of more general deformations was studied in~\cite{SchaferNameki:2010iz}.

The decomposition of IR fields is
\begin{eqnarray}\label{eq:paramPhiU}
\Phi&=& \left(\begin{matrix} Y_{2\tilde N_c \times 2 \tilde N_c} & Z^T_{2\tilde N_c \times 2 (N_f-\t N_c)} \\ - Z_{2(N_f-\t N_c) \times  2 \tilde N_c} &X_{2(N_f-\t N_c) \times2 (N_f-\t N_c)}\end{matrix} \right)\;,\;\\
\Phi_U&=& \left(\begin{matrix} Y_{U,2\tilde N_c \times 2 \tilde N_c} & Z^T_{U,2\tilde N_c \times 2 (N_f-\t N_c)} \\ - Z_{U,2(N_f-\t N_c) \times 2 \tilde N_c} &X_{U,2(N_f-\t N_c) \times 2 (N_f-\t N_c)}\end{matrix} \right)\;,\; \\
q^T&=&\left( \begin{matrix} \chi_{2\tilde N_c \times 2\tilde N_c} \\ \rho_{2(N_f-\t N_c) \times 2\tilde N_c} \end{matrix}\right)\,.
\end{eqnarray}
The $\rho$-fields couple to the linear combination $h \Phi_H \equiv h_1 \Phi + h_2 \Phi_U$. The $X_U$ component (see notation above) of $\Phi_U$ will then be responsible for SUSY breaking after adding the appropriate linear term, and the $Z_U$'s contained within $\Phi_U$ couple to $\rho$, generating vector-like messengers.

Notice the presence of the orthogonal combination to $\Phi_H$ -- call it $\Phi_L$ -- which does not participate in supersymmetry breaking. This distinction of $\Phi_H$ and $\Phi_L$ leads to phenomenological complications. In particular, it is not possible to identically embed a single generation of Standard Model fermions in each of $\Phi$ and $\Phi_U$, since this would lead to vastly different first- and second-generation soft masses in clear conflict with FCNC constraints.

We may solve this problem, as in \cite{Craig:2009hf}, by ``doubling'' the SM matter content in the mesons $\Phi, \Phi_U$ -- i.e., embedding the first generation sfermions in elements of $X_{U,ij}$ that are different from the matrix elements $X_{ij}$ containing the second generation, so that both generations come from the linear combination $\Phi_H$ and acquire comparable one-loop masses. While this solution appears somewhat contrived, it is fairly natural in the $Sp$ models under consideration. Many of the chiral models classified in \S~\ref{subsec:sols} necessarily contain an even number of ${\bf 10}$'s, so that dimensional hierarchy theories built from them {\it must} have the doubling of fields in $\Phi, \Phi_U$ already.

Likewise, following \cite{Craig:2009hf} we may rid ourselves of the {\it faux}-messenger $Z$ field by adding a spectator or sufficiently large stabilizing mass term for components of $\Phi$; this is because $Z$ is not a key component of the CW potential stabilizing the nonsupersymmetric vacuum, and may be disposed of. Since the size of the messenger index is a key consideration in the models we are building, this is generally a useful exercise.

\subsection{Ten-centered models}

In analogy with \S~\ref{subsec:10c}, let us turn to explicit chiral models with a dimensional hierarchy of Standard Model flavor; once again, the simplest such examples are of the ten-centered type.

\subsubsection*{A model of $Sp-1$}
Let's begin with a dimensional hierarchy model of $Sp-1$. The model consists of an  $Sp(2 N_c = 18)$ gauge theory with $N_f = 7$ flavors of fundamentals and embedding 
\begin{center}
\begin{tabular}{c|cc}
&$Sp(18)$&$SU(5)_{SM}$\\
\hline
&&\\[-12pt]
$Q_i^\alpha$& $\Box$ & $({\bf 5 + \bar 5 + 1 + 1}) + {\bf 1 + 1}$  \nonumber\\
$U_{\alpha \beta}$& antisym & $\bf 1$  \nonumber\\
$S_a$& $\bf 1$ & $2 \times \overline{\bf 15} + 2 \times \overline{\bf 10} + 12 \times \overline{\bf 5}$
\end{tabular}
\end{center}
Recall that the anomaly free SM elementary fields are not shown in the spectators above.

The dual gauge theory -- here and in all the other models in this section -- has $\t N_c=2(N_f-2)-N_c = 1$. The structure of the magnetic dual is
\begin{center}
\begin{tabular}{c|cc}
&$Sp(2)$&$SU(5)_{SM}$\\
\hline
&&\\[-12pt]
$M_{ij}$& $\bf 1$ & $8 \times {\bf 5} + 3 \times {\bf 10} + {\bf 15} + 6 \times {\bf 1}$  \nonumber\\
$(M_U)_{ij}$& $\bf 1$ & $8 \times {\bf 5} + 3 \times {\bf 10} + {\bf 15} + 6 \times {\bf 1}$  \nonumber\\
$q_i^\alpha$& $\Box$ & $({\bf 5 + \bar 5} + {\bf 1 + 1}) + {\bf 1 + 1}$  \nonumber\\
$S_a$& $\bf 1$ & $2 \times \overline{\bf 15} + 2 \times \overline{\bf 10} + 12 \times \overline{\bf 5}$
\end{tabular}
\end{center}
The composite messengers are (given that the components of $Z$ are lifted)
\begin{equation}
(\rho \oplus Z_U) \sim 2 \times ( 2 \times ({\bf 5} + {\bf \bar 5}))
\end{equation}
and the lower block of the mesons transform as a $12 \times 12$ antisymmetric tensor, decomposing under $SU(5)$ as
\begin{eqnarray}
X &\sim& 2 \times {\bf{10}} + \left[ {\bf 15 + 10} + 4 \times {\bf 5} + {\bf 1} \right] \nonumber \\
X_U &\sim& 2 \times {\bf{10}} + \left[ {\bf 15 + 10} + 4 \times {\bf 5} + {\bf 1} \right]
\end{eqnarray}
Both $X$ and $X_U$ contain a pair of ${\bf 10}$ fields, so that this model naturally gives us the doubling of fields required for a realistic soft spectrum. Provided we embed the first and second generations in different ${\bf 10}$'s in each meson, both generations of SM sfermions will reside in $\Phi_H$ and get masses at one loop from the CW potential. The messenger index for this theory is $N_{mess} = 4$, nicely compatible with perturbative gauge coupling unification.

\subsubsection*{A model of $Sp-2$}

Consider now the $Sp-2$ embedding with $Sp(2 N_c = 18)$ gauge theory with $N_f = 7$ flavors of fundamentals and customary embedding 
\begin{center}
\begin{tabular}{c|cc}
&$Sp(18)$&$SU(5)_{SM}$\\
\hline
$Q_i^\alpha$& $\Box$ & $({\bf 10} +{\bf 1} + {\bf 1})+ {\bf 1} +{\bf 1}$  \nonumber\\
$U_{\alpha \beta}$& antisym & $\bf 1$  \nonumber\\
$S_a$& $\bf 1$ & $2 \times (\bf 45 + 2 \times \overline{\bf 10})$
\end{tabular}
\end{center}
The structure of the magnetic dual is
\begin{center}
\begin{tabular}{c|cc}
&$Sp(2)$&$SU(5)_{SM}$\\
\hline
&&\\[-12pt]
$M_{ij}$& $\bf 1$ & $ 4 \times {\bf 10} + {\bf \overline{45}} + 6 \times {\bf 1}$  \nonumber\\
$(M_U)_{ij}$& $\bf 1$ & $ 4 \times {\bf 10} + {\bf \overline{45}} + 6 \times {\bf 1}$  \nonumber\\
$q_i^\alpha$& $\Box$ & $(\overline{\bf 10} +{\bf 1} + {\bf 1})+ {\bf 1} +{\bf 1}$  \nonumber\\
$S_a$& $\bf 1$ & $2 \times (\bf 45 + 2 \times \overline{\bf 10})$
\end{tabular}
\end{center}

The composite messengers are
\begin{equation}
(\rho \oplus Z_U) \sim 2 \times ({\bf 10} + {\bf {\overline {10}}}+ 4 \times {\bf 1})
\end{equation}
and the lower block of the mesons transform as a $12 \times 12$ antisymmetric tensor, decomposing under $SU(5)$ as
\begin{eqnarray}
X &\sim& 2 \times {\bf{10}} + \left[{\bf{\overline{45}}} + {\bf 1} \right] \nonumber \\
X_U &\sim& 2 \times {\bf{10}} + \left[{\bf{\overline{45}}} + {\bf 1} \right] \
\end{eqnarray}
As in the previous case, both $X$ and $X_U$ contain two ${\bf 10}$ fields.  Given this matter content, the messenger index for this theory is $N_{mess} = 6$, just compatible with perturbative gauge coupling unification.

\subsubsection*{A model of $Sp-3$}

Last among the ten-centered models, consider the $Sp-3$ embedding with $Sp(2 N_c = 26)$ gauge theory with $N_f = 9$ flavors of fundamentals and embedding 
\begin{center}
\begin{tabular}{c|cc}
&$Sp(18)$&$SU(5)_{SM}$\\
\hline
&&\\[-12pt]
$Q_i^\alpha$& $\Box$ & $ ({\bf 5} + {\bf 5} + {\bf \bar 5} + {\bf 1})+ {\bf 1} +{\bf 1}$  \nonumber\\
$U_{\alpha \beta}$& antisym & $\bf 1$  \nonumber\\
$S_a$& $\bf 1$ & $2 \times \overline{\bf 10} + 2 \times \overline{\bf 15} + 4 \times \overline{\bf 5}$
\end{tabular}
\end{center}

The structure of the magnetic dual is
\begin{center}
\begin{tabular}{c|cc}
&$Sp(2 \t N_c)$&$SU(5)_{SM}$\\
\hline
&&\\[-12pt]
$M_{ij}$& $\bf 1$ & $6 \times {\bf 5} + 3 \times {\bf \bar 5} + 3 \times {\bf 10} + \overline{\bf 10} + {\bf 15} + 2 \times {\bf 24}  + 5 \times {\bf 1}$  \nonumber\\
$(M_U)_{ij}$& $\bf 1$ & $6 \times {\bf 5} + 3 \times {\bf \bar 5} + 3 \times {\bf 10} + \overline{\bf 10} + {\bf 15} + 2 \times {\bf 24}  + 5 \times {\bf 1}$  \nonumber\\
$q_i^\alpha$& $\Box$ & $ ({\bf 5} + {\bf \bar 5} + {\bf \bar 5} + {\bf 1})+ {\bf 1} +{\bf 1}$  \nonumber\\
$S_a$& $\bf 1$ & $2 \times \overline{\bf 10} + 2 \times \overline{\bf 15} + 4 \times \overline{\bf 5}$
\end{tabular}
\end{center}

For the model considered here, we have composite messengers
\begin{equation}
(\rho \oplus Z_U) \sim 2 \times (3 \times {\bf 5} + 3 \times {\bf \bar 5} + 2 \times {\bf 1})
\end{equation}
and the lower block of the mesons transform as a $16 \times 16$ antisymmetric tensor, decomposing under $SU(5)$ as
\begin{eqnarray}
X &\sim&2 \times {\bf{10}} + \left[({\bf 10 + \overline{10}}) + ({\bf 5 + \bar 5}) + {\bf 5} + {\bf 15} + 2 \times {\bf 24}  + 2 \times {\bf 1} \right]\\
X_U &\sim& 2 \times {\bf{10}} + \left[({\bf 10 + \overline{10}}) + ({\bf 5 + \bar 5}) + {\bf 5} + {\bf 15} + 2 \times {\bf 24}  + 2 \times {\bf 1} \right]
\end{eqnarray}
Once again, both $X$ and $X_U$ contain two ${\bf 10}$ fields. Given this matter content, the messenger index for this theory is $N_{mess} = 6$, just compatible with perturbative gauge coupling unification.

\subsection{Model without spectators}\label{subsec:groupth}

Last but not least, we may construct a model of dimensional hierarchy without spectators. The electric theory is
\begin{center}
\begin{tabular}{c|cc}
&$Sp(26)$&$SU(5)_{SM}$\\
\hline
&&\\[-12pt]
$Q_i^\alpha$& $\Box$ & $({\bf 10 + \bar 5 + 1}) + {\bf 1 + 1}$  \nonumber\\
$U_{\alpha \beta}$& antisym & $\bf 1$  \nonumber\\
$S_a$& $\bf 1$ & $-$
\end{tabular}
\end{center}
with magnetic dual
\begin{center}
\begin{tabular}{c|cc}
&$Sp(2)$&$SU(5)_{SM}$\\
\hline
&&\\[-12pt]
$M_{ij}$& $\bf 1$ & $3 \times ( {\bf{10}} + {\bf {\bar 5}}) + \left[{\bf 45}+{\bf{\overline{45}}}+ {\bf{\overline{10}}}+ {\bf 5}+ 3 \times {\bf 1} \right]$  \nonumber\\
&&\\[-12pt]
$(M_U)_{ij}$& $\bf 1$ & $3 \times ( {\bf{10}} + {\bf {\bar 5}}) + \left[{\bf 45}+{\bf{\overline{45}}}+ {\bf{\overline{10}}}+ {\bf 5}+ 3 \times {\bf 1} \right]$  \nonumber\\
$q_i^\alpha$& $\Box$ & $({\bf \overline{10} + 5 + 1}) + {\bf 1 + 1}$  \nonumber\\
$S_a$& $\bf 1$ & $-$
\end{tabular}
\end{center}
In order to ensure the correct Standard Model index at low energies and remove the superfluous charged matter contained in $Z$, we may lift the excess $\bf{10} + {\bf {\bar 5}}$ vector-like pairs in $Z, X, X_U$ through unusual but technically natural superpotential terms of the form
\begin{equation}\label{eqn:weirdlift}
\Delta W = Z_{(10 + \bar 5)} X_{(\overline{10} + 5)} + Z_{(10 + \bar 5)'} X_{U, (\overline{10} + 5)}
\end{equation}
These would correspond to dimension 4 and higher in the electric theory. The composite messengers are then
\begin{equation}
(\rho \oplus Z_U) \sim 2 \times ( {\bf 10 + \overline{10}} + {\bf \bar 5 + 5} + 2 \times {\bf 1})
\end{equation}
and the lower block of the mesons transform as a $16 \times 16$ antisymmetric tensor, decomposing under $SU(5)$ as
\begin{eqnarray}
X & \sim & ( {\bf{10}} + {\bf {\bar 5}}) + \left[{\bf 45}+{\bf{\overline{45}}}+ {\bf{\overline{10}}}+ {\bf 5} \right] \nonumber\\
X_U & \sim & ( {\bf{10}} + {\bf {\bar 5}}) + \left[{\bf 45}+{\bf{\overline{45}}}+ {\bf{\overline{10}}}+ {\bf 5} \right] \,.
\end{eqnarray}
The excess $({\bf{\overline{10}}}+ {\bf 5})$'s in each of $X, X_U$ may be removed without spectators by the deformation (\ref{eqn:weirdlift}), while the $({\bf 45}+{\bf{\overline{45}}})$ may be given a vector mass. The appeal of such a construction rests in its utter lack of spectators, but there are a variety of shortcomings. Already the theory has a messenger index $N_{mess} = 8$, leading to a Landau pole well shy of the GUT scale. Perturbative unification aside, there are a variety of other shortcomings; among other things the pseudomodulus $X_U$ lacks a requisite Standard Model gauge singlet, while the embedding of a single ${\bf 10 + \bar 5}$ in each of $X, X_U$ leads to a phenomenologically unviable soft spectrum for reasons discussed earlier.

In order to rectify these shortcomings, the most minimal extension of this theory requires
$$N_f = 10, \; \; N_c = 15, \; \; n_1 = 5$$
The resulting spectrum is a natural generalization of the above, but now possesses both an SM gauge singlet in $X_U$ and the desired $({\bf 10 + \bar 5})$ `doubling' in each of $X, X_U$ required for a viable soft spectrum. However, this doubling leads to extra light $({\bf 10 + \bar 5})$ fermions which may only be removed by spectators. Hence this phenomenologically-viable dimensional hierarchy model no longer strictly qualifies as ``spectator-free''.

\section{Conclusions}

Single-sector models offer an exceptionally compact means of realizing a variety of beyond-the-Standard Model objectives: explaining the origin of quarks and leptons, generating the flavor hierarchy observed in Yukawa matrices, breaking supersymmetry dynamically, and naturally communicating its breaking in a flavor-respecting manner to Standard Model superfields without the need for a separate messenger sector. However, the majority of single-sector models constructed to date have suffered from a range of shortcomings, including an excess of SM exotics, Landau poles in the gauge couplings below the unification scale and no dynamical explanation for the Standard Model index. Additional spectators were required to lift unwanted states, leading in some cases to unwanted (and phenomenologically unviable) new vacua. 

In this work we have attempted to address these shortcomings. To this end, we have found a small number of models based on $Sp(N)$ SSQCD in which supersymmetry is broken in a metastable vacuum and the Standard Model index is generated dynamically. The nontrivial family structure is explained in terms of electric ``preons'' that transform under chiral representations of the SM. These models are quite compact, with few added spectators and ready compatibility with perturbative unification of gauge couplings. In $Sp(N)$ theories with only fundamental matter, the single-sector models realize a $U(2)$ flavor symmetry that explains the heaviness of the top quark and leads to a soft spectrum free of problematic FCNCs, but must be supplemented by additional flavor textures to explain the pattern of light fermion masses. In contrast, $Sp(N)$ theories with both fundamental and antisymmetric matter realize a dimensional hierarchy of fermion masses, which is alone sufficient to explain the observed Yukawa textures but is free from FCNCs only over a particular range of parameters. 

In either case, the resulting soft spectrum offers a natural ultraviolet realization of the ``more minimal'' soft spectrum with a gravitino LSP and direct mediation. Both types of theories may be used to construct a variety of specific models, depending on the way in which Standard Model representations are embedded in the global flavor symmetry. In most cases, such compact models realize a ten-centered flavor texture with a small number of additional spectators whose existence is often required by anomaly cancellation. There is, however, also a unique model with {\it no spectators} in which both ${\bf 10}$ and ${\bf \bar 5}$ are composite, though the messenger index of this theory is incompatible with perturbative unification. 

There has often been a tension between the pleasing conciseness of the single-sector principle and the variety of actual single-sector realizations. While the models constructed here are far from perfect, we hope that they represent progress in developing more compact, calculable models that naturally combine flavor and supersymmetry breaking in four dimensions. As these models grow more realistic, it would be useful to determine how their structure may be probed more deeply at colliders. Certainly evidence for a ``more minimal'' spectrum of soft masses at the LHC with gravitino LSP would be a compelling first indication. But it may also be the case that experimental indications first arise in other settings; in particular, there is the intriguing possibility for signals of single-sector physics to arise in the $B_s$ meson system, along the lines of \cite{Giudice:2008uk}.

Finally, it is quite interesting that these single-sector models admit a systematic classification, which moreover reveals a dynamical relation between chirality and unification. How general is this classification? It certainly does not include models where the subgroup that confines is itself chiral, or where there is no product gauge group structure at all. Both alternatives deserve being explored in depth and may reveal a very rich dynamics --but it is also much harder to explore them systematically. On the other hand, our classification is general granting a product gauge group structure and that the strong dynamics is vector-like. The analysis for SU and SO groups is very similar to the Sp case considered in this work (exceptional groups are not included here). Similarly, the addition of other 2-index representations is expected to parallel the antisymmetric case. Representations with more indices are quickly forbidden by asymptotic freedom.

\section*{Acknowledgments}
We are grateful to R. Essig, S. Franco, A. Hook, S. Kachru, A. Nacif, S. Sch\"afer-Nameki and C. Tamarit for helpful discussions and useful comments on the draft. We would like to thank D. Green, A. Katz, and Z. Komargodski for discussions regarding the vacuum structure of SQCD deformed by gauge singlets. We especially thank D. Green for extensive conversations related to chiral metastable vacua. NC thanks the Dalitz Institute for Fundamental Physics and the Department of Theoretical Physics, Oxford University for hospitality. GT thanks the KITP for hospitality. NC is supported by the NSF GRFP and NSF Grants 0756174 and 0907744. G.T. is supported by the US DOE under contract number DE-AC02-76SF00515 at SLAC.

\appendix

\section{Limits on the  messenger index}\label{sec:unif}

\begin{figure}[] 
  \centering
      \includegraphics[width=2.5in]{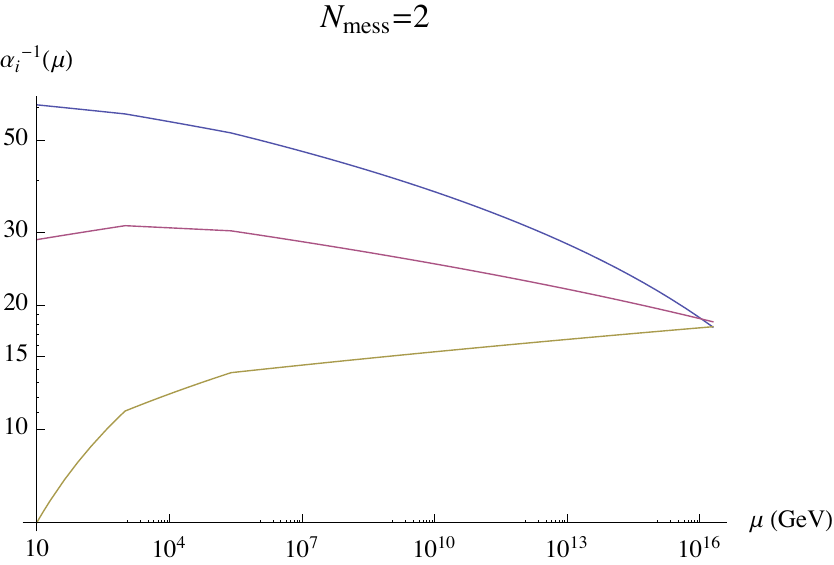} 
         \includegraphics[width=2.5in]{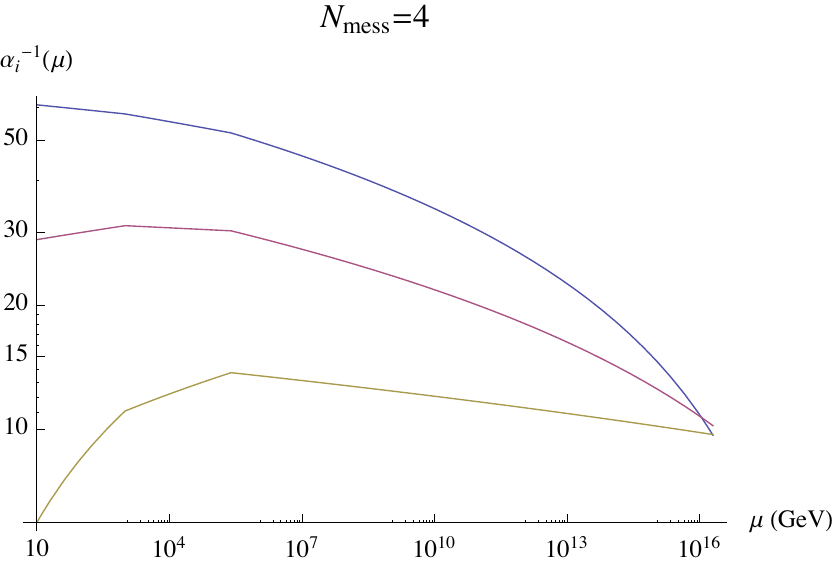} 
            \includegraphics[width=2.5in]{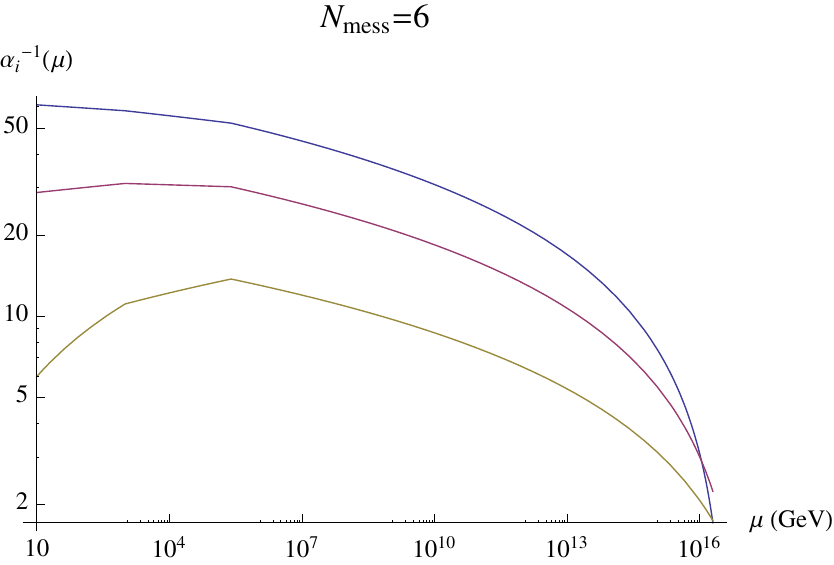} 
   \includegraphics[width=2.5in]{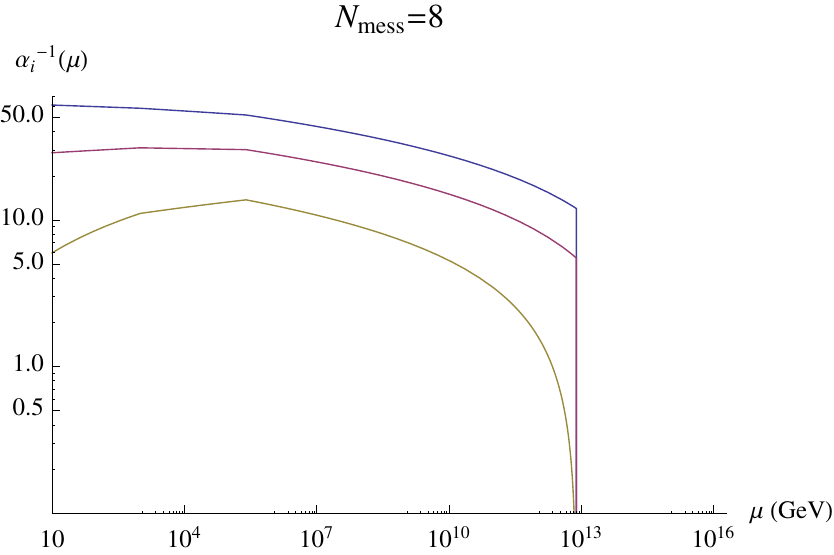} 
      \caption{One-loop running of Standard Model gauge couplings for different values of the messenger index $N_{mess}$ given sparticle masses $\tilde m \sim 1 - 10$ TeV and messenger masses $M_{mess} \sim 250$ TeV. Perturbative unification allows $N_{mess} \leq 6$; $N_{mess} \leq 4$ is required if the messengers are significantly lighter ($M_{mess} \lesssim 100$ TeV).}
   \label{fig:nmess}

\end{figure}

Here we briefly review the constraints on the messenger index from perturbative unification of Standard Model gauge couplings. For simplicity, we merely compute the one-loop running of SM gauge couplings between the weak scale and unification scale, taking into account thresholds associated with the appearance of messenger matter; the results are shown in Fig.~\ref{fig:nmess}. We take messenger masses to lie around 250 TeV (the natural value in single-sector models). In this case, $N_{mess} \leq 6$ is readily compatible with perturbative gauge coupling unification. In contrast, $N_{mess} = 8$ is strongly disfavored with a Landau pole at $\mu \sim 10^{12}$ GeV. From the perspective of flavor model-building it is perhaps unnecessary to require perturbative unification of SM gauge couplings, but perturbativity is nonetheless a necessary ingredient if we are to retain calculability in the models under consideration.

\section{FCNCs in single-sector models}\label{sec:FCNC}

There are two primary contributions to FCNCs in the models under consideration: (1) contributions coming from direct coupling between SM fermions and supersymmetry-breaking messengers; and (2) conventional gluino-mediated contributions involving MSSM sparticles due to the misalignment of scalar soft masses and fermion masses. The former do not lead to significant constraints on the low-energy spectrum, as box diagrams involving messengers are suppressed both by a loop factor and the large messenger mass; the most stringent constraints coming from $K^0 - \bar K^0$ mixing are readily satisfied by $\mu \gtrsim 160$ TeV \cite{Craig:2009hf}.

However, the latter contributions from MSSM sparticles do constrain the weak-scale soft spectrum, particularly for dimensional hierarchy models. We will parametrize the gluino-mediated contributions to 
flavor changing neutral currents (FCNCs) following \cite{Gabbiani:1996hi}; our notational conventions are those of \cite{Craig:2009hf}. We may place bounds on first- and second-generation sfermion masses for both democratic and ten-centered single-sector flavor models from $K^0 - \bar{K}^0$, $D^0 - \bar{D}^0$, or $B^0 - \bar B^0$ mixing 
and the rare decays $\mu\to e\gamma$ and $b \to s \gamma.$  Naturalness dictates that the stop mass lie around 1-2 TeV, which sets the scale of gauge-mediated contributions to all three generations.  When this is the only source of SUSY breaking, SUSY FCNCs are negligible. However, in addition to the gauge-mediated contribution, the first and second generation 
squarks and sleptons may obtain additional soft masses directly from SUSY-breaking, leading to an inverse hierarchy. The size of additional contributions to the soft masses $m_{\tilde f_1}, m_{\tilde f_2}$ of the first two generations is then constrained by FCNCs.

In the case of $U(2)$ symmetric models, all FCNCs involving first- and second-generation sfermions are naturally mitigated, as the soft masses for these generations are universal. Principal constraints then arise from FCNC's involving first- and third-generation sfermions, e.g., from $B^0 - \bar B^0$ mixing. Since limits on $B^0 - \bar B^0$ mixing are much more relaxed than those on $K^0 - \bar K^0$ mixing, the soft spectrum in $U(2)$ symmetric models is essentially unconstrained by flavor considerations. In the case of dimensional hierarchy models, the primary constraints still arise from typical processes involving first- and second-generation sfermions -- primarily $K^0 - \bar K^0$ in the case of democratic models and $D^0 - \bar D^0$ in the case of ten-centered models (where the right-handed contributions to $K^0 - \bar K^0$ vanish). As the yukawa textures for democratic and ten-centered models differ significantly, we will consider the relevant constraints in turn.

\subsection{Constraints}

Straightforward constraints on sparticle masses from FCNCs may be placed by considering gluino-mediated contributions to neutral meson mixing and decay. The Standard Model contributions to measured meson mixings fall within the measured values, but depend on hadronic uncertainties to an extent that the full contribution is unknown. Thus we may take as our constraints the requirement that contributions to $\Delta m_K, \Delta m_D, \Delta m_B,$ and $\Gamma(b \to s \gamma)$ do not exceed (in magnitude) the measured values.  We extract the contribution to these quantities from squark mixing from \cite{Gabbiani:1996hi}. The relevant processes are 

\begin{itemize}

\item $K^0 - \bar K^0$: We may constrain the possible values of $m_{\tilde f_1}$ and $m_{\tilde f_2}$ via the parameters $(\delta^d_{LL})_{12}$ and $(\delta^d_{RR})_{12}$, by computing their contribution to the $K_L - K_S$ mass difference $\Delta m_K.$ This difference has been measured within excellent precision to be very nearly $\Delta m_K = (3.483 \pm 0.006) \times 10^{-12}$ MeV \cite{Amsler:2008zzb}.  These contributions depend on the gluino mass $m_{\tilde g}$ and the squark masses $m_{\tilde f_1}, m_{\tilde f_2}$.
 
\item $D^0 - \bar D^0$: Similar constraints on $(\delta^u)_{12}$ arise from $D^0 - \bar{D}^0$ mixing via their contribution to 
 $\Delta m_D = (1.57^{+.39}_{-.41} ) \times 10^{-11}$ MeV  \cite{Amsler:2008zzb}. 

\item $B^0 - \bar B^0$: The mixings $(\delta^d_{MN})_{13}$ may similarly be constrained by $B^0 - \bar{B}^0$ mixing from their contribution to 
$\Delta m_B = (3.337 \pm 0.033) \times 10^{-10}$ MeV \cite{Amsler:2008zzb}. 

\item $b \to s \gamma$ and $\mu \to e \gamma$: We may constrain mixing between the second and third generations via the rare decay $b \to s \gamma$, using the gluino-mediated 
contribution in  \cite{Gabbiani:1996hi}. In this case, we require that our contribution not exceed the measured branching ratio $BR(b \to s 
\gamma) = (3.52 \pm 0.23 \pm 0.09) \times 10^{-4}$ \cite{Barberio:2008fa}. The branching ratio is a strong function of squark mass, and is satisfied readily for squark masses above 1 TeV; thus $b \to s \gamma$ does not place a significant constraint on the models under consideration. Similar constraints may be placed on the lepton sector via the rare decay $\mu \to e \gamma$, but once again these limits are readily satisfied by the sparticle masses under consideration. Constraints from $b \to s \gamma$ and $\mu \to e \gamma$ will therefore not be shown in what follows.
\end{itemize}

\subsubsection*{Constraints from tachyonic stop mass}

We must also take into account an upper bound placed on squark masses by the requirement of a positive stop mass at the weak scale. As 
noted in \cite{AHM}, overly large masses for the first and second-generation squarks can drive the stop mass negative via their two-loop 
contribution to the stop mass RG.  This places an upper bound on soft masses for the first two generations of squarks; we require merely that the stop retain a positive mass-squared at the weak scale. The constraints relevant to our models were studied in~\cite{Craig:2009hf}; we refer the reader to their Appendix A.5 for details. With these constraints in hand, let us now turn to explicit limits on the soft spectrum.



\subsection{Models}

Now we may turn to detailed limits on the soft spectrum for the four classes of yukawa textures under consideration. The constraints on democratic models are much the same as those appearing in \cite{Franco:2009wf, Craig:2009hf}, which results in fairly stringent limits on the first two generations. In contrast, ten-centered models offer an exceptionally appealing scenario from the perspective of FCNCs, in that the strongest constraints from $K^0 - \bar K^0$ mixing only arise from the $LL$ sector and are naturally mitigated.  In GUT models where only the $Q, \bar u, \bar e \in {\bf 10}$ of the first two generations receive SUSY-breaking soft masses directly from $V_{CW}$, the remaining fields obtain strictly flavor-universal masses from gauge mediation. Crucially, this spectrum still guarantees a vanishing hypercharge $D$-term at leading order, preserving the successes of the ``more minimal'' spectrum even though not all first- and second-generation sfermions are heavy. In detail:

\begin{figure}[h!]
   \centering
   \includegraphics[width=3in]{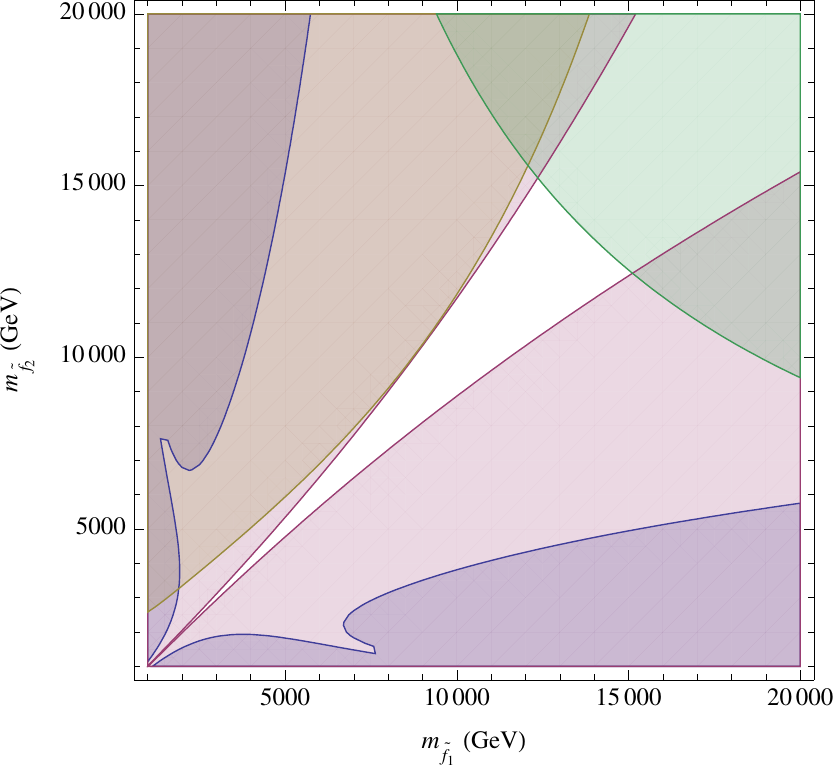} 
      \includegraphics[width=3in]{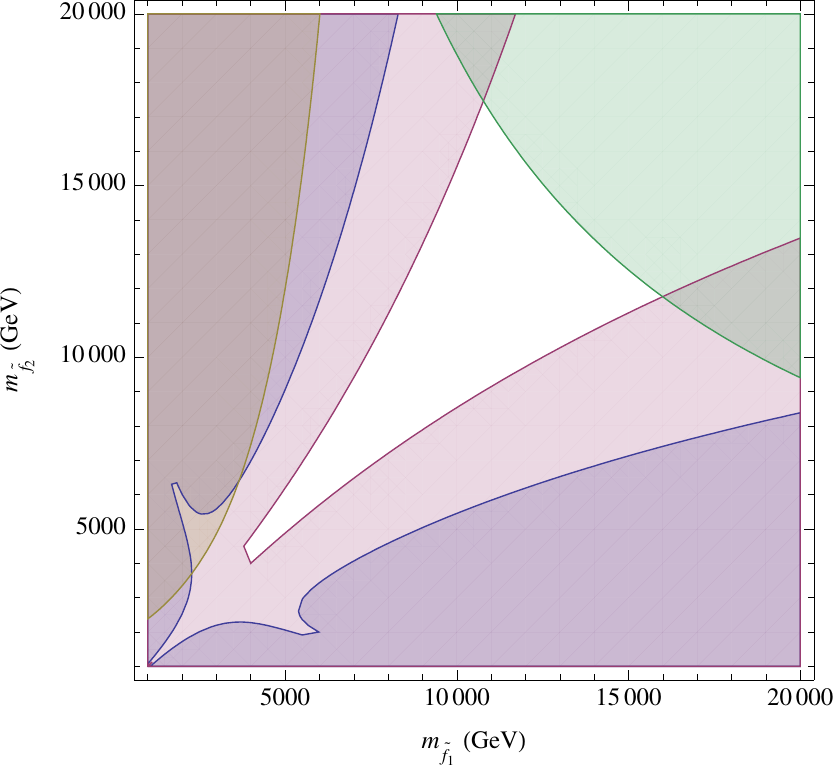} 
      \includegraphics[width=3in]{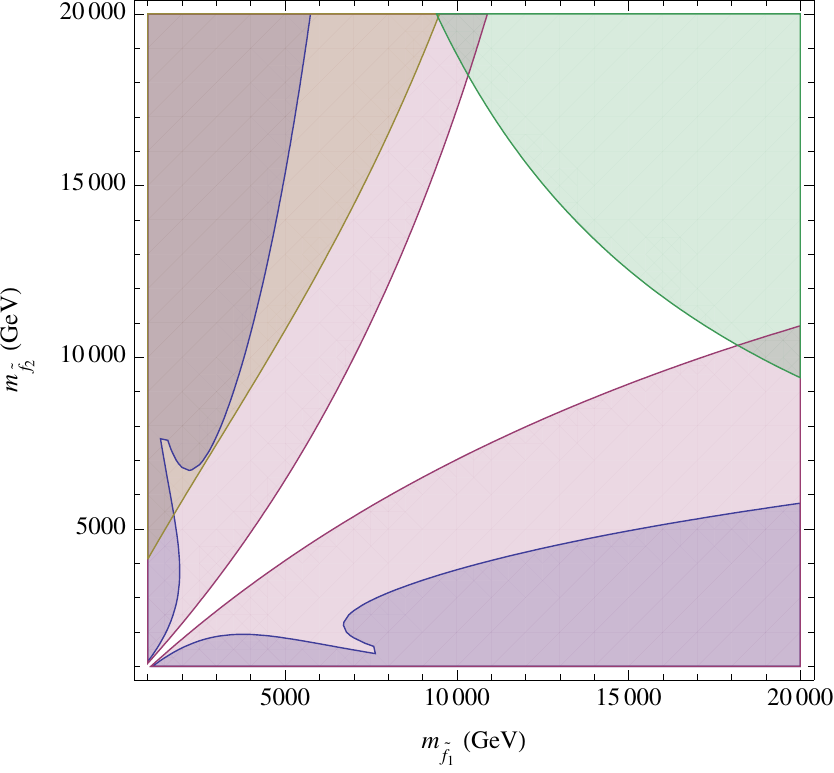} 
   \includegraphics[width=3in]{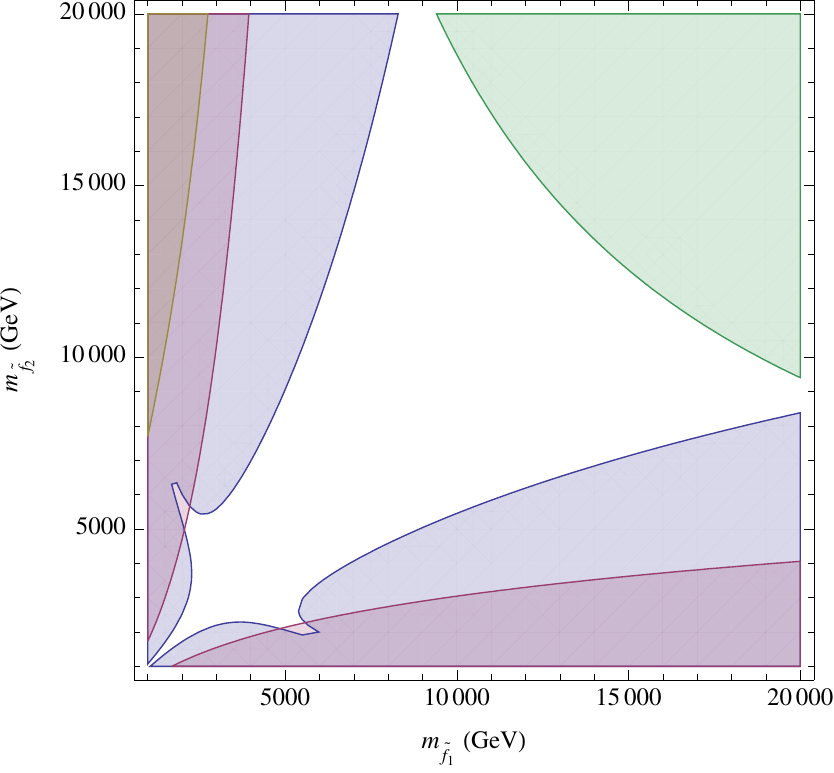} 

   \caption{FCNC constraints on first- and second-generation soft masses for (a) democratic, $U(2)$-symmetric models (upper left), (b) democratic, dimensional hierarchy models (upper right), (c) ten-centered, $U(2)$-symmetric models (lower left), and (d) ten-centered, dimensional hierarchy models (lower right) given $m_{\tilde f_3} = 1$ TeV and $m_{\tilde g} = 500$ GeV. The purple region is excluded by $K^0 - \bar K^0$; the blue region is excluded by $D^0 - \bar D^0$; the brown region is excluded by $B^0 - \bar B^0$.  The green region is excluded by tachyonic stop masses. Note that, for $U(2)$-symmetric models, the first- and second-generation masses are identical, so that FCNC constraints are automatically satisfied.}
   \label{fig:fcncs}
\end{figure}

\begin{itemize}
\item Democratic, $U(2)$-symmetric: The primary constraint in this case comes in principle from $B^0 - \bar B^0$ mixing, as all constraints involving first- and second-generation sfermions are automatically satisfied. Even in this case, however, the $B^0 - \bar B^0$ limits do not place any stringent constraints on the model.  

\item Democratic, dimensional hierarchy: Primary constraints in this case come from $K^0 - \bar K^0$ mixing. The soft spectrum is heavily proscribed by the combination of $K^0 - \bar K^0$ mixing limits and positivity of the stop mass; this is the most constrained of the model types under consideration. 

\item Ten-centered, $U(2)$-symmetric: As with the democratic case, the primary constraint in these models comes from $B^0 - \bar B^0$ mixing. However, in ten-centered models the only contributions to $B^0 - \bar B^0$ mixing come from the $LL$ sector, making the constraints even weaker; consequently the soft spectrum is effectively unconstrained by FCNCs.

\item Ten-centered, dimensional hierarchy: Primary constraints in this case come from $D^0 - \bar D^0$ mixing, as limits from $K^0 - \bar K^0$ mixing are weakened  since only the $LL$ sector contributes.
\end{itemize}

Of course, in order to compute bounds from FCNCs it is necessary to begin with Yukawa matrices that reproduce Standard Model fermion masses and CKM mixing angles as accurately as possible. As a numerical example for flavor textures for $U(2)$-symmetric theories, we consider
$$
v_1 =4\times10^{-1}\;,\;\epsilon=10^{-1}\;,\;v_2=4\;,\;\tan\,\beta=20\,.
$$
For a `democratic' $U(2)$ model we may then choose
Yukawa matrices
\begin{equation}
Y_{u} = \left( \begin{array}{ccc}
0 & 0.3 \,v_1 \epsilon^3 & 0 \\
0.3\,v_1\epsilon^3 &3\,v_2^2\epsilon^4  & v_2 \epsilon^2    \\
0 &v_2 \epsilon^2  & 0.8
\end{array}
\right)\;\;,\;\;
Y_{d} = \left( \begin{array}{ccc}
0 & 3.2\,v_1\epsilon^3 & 0 \\
3.2\,v_1\epsilon^3 &6\,v_2^2\epsilon^4 & 0.9 \,v_2 \epsilon^2  \\
0 &0.9\,v_2 \epsilon^2 & 0.3
\end{array}\right)\,.
\end{equation}
This gives a top mass of order $150$ GeV, and a bottom mass $\sim 2.5$ GeV at the TeV scale.

It is interesting that with the same set of parameters, we can also have a realistic ten-centered Yukawa matrix
\begin{equation}
Y_{d} \sim \left( \begin{array}{ccc}
0 & 0.3\,v_1\epsilon^2 & 0 \\
0.3\,v_1\epsilon^2 &0.6\,v_2^2\epsilon ^3 &0.4\,v_2 \epsilon^2 \\
0 &0.2  \, v_2 \epsilon  & 0.3
\end{array}
\right)~.
\end{equation} 
Notice that, with these Yukawas, the CKM parameters are set by the ratio of quark masses.
For instance, the Cabibbo angle is $\sqrt\frac{m_d}{m_s}\simeq 0.2$~\cite{flavorU2}.

For dimensional hierarchy models, the additional texture makes it unnecessary to rely on additional flavor spurions, though $\mathcal{O}(1)$ numerical coefficients are still required. We use for a `democratic' model the numerical textures
\begin{equation}
Y_{u} \sim \left( \begin{array}{ccc}
0.5 \epsilon^4 & 1.2 \epsilon^3 &  \epsilon^2 \\
1.2 \epsilon^3  & 3 \epsilon^2  & 2 \epsilon     \\
\epsilon^2 &2 \epsilon   & 1
\end{array}
\right) \; \; , \; \;
Y_{d} \sim \left( \begin{array}{ccc}
2 \epsilon^4 & 4 \epsilon^3 & 0.5 \epsilon^2 \\
4 \epsilon^3  & 6 \epsilon^2  & 2 \epsilon     \\
0.5 \epsilon^2 &2 \epsilon   & 0.5
\end{array}
\right)~,
\end{equation}
which reproduce Standard Model fermion masses and the CKM parameters with a fair degree of accuracy. For ten-centered models we take instead the down-type Yukawa texture
\begin{equation}
Y_{d} \sim \left( \begin{array}{ccc}
0.35 \epsilon^2 & 0.5 \epsilon^2  & \epsilon^2\\
0.17 \epsilon  & 0.16 \epsilon  & \epsilon     \\
0.1 &0.25  & 1
\end{array}
\right)~.
\end{equation}
Detailed constraints are shown in Fig. \ref{fig:fcncs}.

\bibliographystyle{JHEP}
\renewcommand{\refname}{Bibliography}
\addcontentsline{toc}{section}{Bibliography}
\providecommand{\href}[2]{#2}\begingroup\raggedright

\end{document}